\documentclass{elsart}
\usepackage{epsfig}
\newcommand{\pme}[2]{{}^{+#1}_{-#2}}
\newcommand{\Bhatt}{Bhattacharya \emph{et al.} }
\newcommand{\order}[1]{{\mathcal O}(#1)}
\newcommand{\cl}[1]{\mathcal{#1}}
\newcommand{\psb}{\bar{\psi}}
\newcommand{\middle}[1]{\raisebox{1.5ex}[0pt]{#1}}

\makeatletter
\def\section{\@startsection{section}{1}{\z@}{1.2\@bls
  \@plus .4\@bls \@minus .1\@bls}{0.3\@bls}{\normalsize\bfseries}}
\def\subsection{\@startsection{subsection}{2}{\z@}{0.8\@bls
  \@plus .3\@bls \@minus .1\@bls}{0.05\@bls}{\normalsize\itshape}}
\def\subsubsection{\@startsection{subsubsection}{3}{\z@}{0.5\@bls
  \@plus .2\@bls}{0.0001pt}{\normalsize\itshape}}
\def\paragraph{\@startsection{paragraph}{4}{\z@}{3.25ex \@plus
  2ex \@minus 0.2ex}{-1em}{\normalsize\bfseries}}
\floatsep \@bls \@plus 4\p@ \@minus 2\p@ 
\def\eqnarray{%
  \par                                               
  \noindent                                          
  \baselineskip\eqnbaselineskip\lineskip\eqnlineskip 
  \lineskiplimit\eqnlineskip                         
  \stepcounter{equation}%
  \let\@currentlabel=\theequation
  \global\@eqnswtrue
  \global\@eqcnt\z@
  \tabskip\mathindent
  \let\\=\@eqncr
  \abovedisplayskip\eqntopsep\ifvmode\advance\abovedisplayskip\partopsep\fi
  \belowdisplayskip\eqntopsep
  \belowdisplayshortskip\abovedisplayskip
  \abovedisplayshortskip\abovedisplayskip
  $$\halign to \displaywidth\bgroup\@eqnsel
    \pre@coli$\displaystyle\tabskip\z@{##}$\post@coli
    &\global\@eqcnt\@ne
    \pre@colii$\displaystyle{##}$\post@colii
    &\global\@eqcnt\tw@
    \pre@coliii $\displaystyle\tabskip\z@{##}$\post@coliii
    \tabskip\@centering&\llap{##}\tabskip\z@\cr
}
\makeatother

\begin{document}
\begin{flushright}\begin{small}
Edinburgh 2000/14\\
IFUP-TH/2000-17 \\
JLAB-THY-00-25 \\
SHEP 00 08
\end{small}\end{flushright}
\vskip1ex
\begin{frontmatter}
\title{Decay Constants of B and D Mesons from Non-perturbatively 
        Improved Lattice QCD}
\collab{UKQCD Collaboration}
\author[ed]{K.C.~Bowler}, 
\author[soton]{L.~Del~Debbio\thanksref{pisa}},
\author[soton]{J.M.~Flynn},
\author[ed]{G.N.~Lacagnina},
\author[soton]{V.I.~Lesk\thanksref{tsukuba}},
\author[ed]{C.M.~Maynard},
\author[ed,jefferson,ODU]{D.G.~Richards}
\thanks[pisa]{Dipartimento di Fisica, Universit\`{a} di Pisa and 
        INFN Sezione di Pisa, Italy.} 
\thanks[tsukuba]{Center for Computational Physics, Tsukuba Daigaku,
        Tsukuba Ibaraki, 305-8577, Japan}
\address[ed]{Department of Physics \& Astronomy, University of Edinburgh,
        Edinburgh EH9 3JZ, Scotland, UK}
\address[soton]{Department of Physics \& Astronomy, University of Southampton,
        Southampton, SO17 1BJ, UK}
\address[jefferson]{Jefferson
        Laboratory, MS 12H2, 12000 Jefferson Avenue, Newport News, VA 23606,
        USA.}
\address[ODU]{Department of Physics, Old Dominion University, Norfolk, 
        VA 23529, USA.}
\date{\today}

\begin{abstract}
The decay constants of B, D and K mesons are computed in quenched lattice
QCD at two different values of the coupling. The action and operators
are $\mathcal{O}(a)$ improved with non-perturbative coefficients. 
The results are $f_{\mathrm B}=195(6)\pme{24}{23}$ MeV,
$f_{\mathrm D}=206(4)\pme{17}{10}$ MeV,
$f_{\mathrm{B_s}}=220(6)\pme{23}{28}$ MeV,
$f_{\mathrm{D_s}}=229(3)\pme{23}{12}$ MeV and 
$f_\mathrm{K}=150(3)\pme{12}{\ 8}$ MeV.
Systematic errors are discussed in detail. Results for vector decay
constants, flavour symmetry breaking ratios of decay constants, the
pseudoscalar-vector mass splitting and D meson masses are also
presented.
\end{abstract}
\end{frontmatter}


\section{Introduction}
The accurate determination of the B and D meson decay constants is
of profound importance in phenomenology.  The combination $f_{\mathrm
B}\sqrt{ B_{\mathrm B}}$, for both $\mathrm B_{\mathrm d}$ and
$\mathrm B_{\mathrm s}$ mesons, plays a crucial role in the extraction
from experimental data of CKM quark mixing and CP violation
parameters.  The phenomenological parameter $B_{\mathrm{B}}$ describes
$\mathrm B^0$--$\bar \mathrm B^0$ mixing, and is expected to be close
to unity.  D meson decay constants are needed for calculations based
on factorisation of non-leptonic B meson decays to charmed mesons
(see~\cite{neustech} for a review and~\cite{ccfm} for a recent
application).

Numerical simulations of lattice QCD provide a method for computing
the requisite matrix elements from first principles. A prime concern
in such calculations is the control of discretisation errors, most
notably those associated with heavy quarks, given that in practice the
$b$-quark mass is large in lattice units: $am_Q>1$. One
approach, pioneered by Bernard \emph{et al.}~\cite{bernard_prd38} and
by Gavela \emph{et al.}~\cite{gavela_plb206} and used in the present
study, is to work with heavy quarks around the charm quark mass and
extrapolate to the $b$ mass scale guided by continuum heavy-quark
effective theory (HQET). Other techniques use some form of effective
field theory directly to reduce the cut-off effects associated with
heavy quarks. Examples include lattice HQET~\cite{eichten},
non-relativistic QCD (NRQCD)~\cite{lepage_thacker} and a
re-interpretation of the Wilson fermion action as an effective theory
for heavy quarks, developed by El~Khadra \emph{et
al.}~\cite{KLM_norm,kron_prd62}, known as the Fermilab formalism. A
critical discussion of these various methods, their associated
systematic errors and a survey of recent results for $f_{\mathrm{B}}$,
is given in a comprehensive review by Bernard~\cite{Bernard_2000}.

This paper presents decay constants and masses of heavy-light mesons
calculated in the quenched approximation to QCD at two values of the
lattice coupling, $\beta=6.2$ and $\beta=6.0$.  The calculation uses a
non-perturbatively improved relativistic Sheikholeslami-Wohlert
(SW)~\cite{sw_paper} fermion action and current operators so that the
leading discretisation errors in lattice matrix elements appear at
$\order{a^2}$ rather than $\order{a}$. However, non-perturbative
improvement does not necessarily reduce lattice artefacts at a given
$\beta$. Given results at two $\beta$ values, heavy quark
mass-dependent discretisation errors can be estimated by combining a
continuum extrapolation with the fit in heavy quark mass, although a
simple continuum extrapolation of the final result is not attempted.

Details of the lattice calculation and the extraction of decay
constants and masses from Euclidean Green functions are described in
Section~\ref{sec:details}.  The extrapolations to physical quark
masses, both heavy and light, and the heavy quark symmetry (HQS)
relationship between pseudoscalar and vector decay constants are
discussed in Section~\ref{sec:extrap}.  Results for the decay
constants are presented in Section~\ref{sec:decay_const} and
summarised here.
\begin{center}
\begin{tabular}{ll}
\renewcommand{\arraystretch}{1.4}
$\begin{array}{rcl}
f_\mathrm{B} & = &195(6)^{+24}_{-23}\ \mathrm{MeV} \\ 
f_\mathrm{D}& = & 206(4)^{+17}_{-10}\ \mathrm{MeV}\\
f_\mathrm{B_{S}} &= &220(6)^{+23}_{-28}\ \mathrm{MeV} \\
f_\mathrm{D_{S}}& = & 229(3)^{+23}_{-12}\ \mathrm{MeV} \\[2ex]
f_{\mathrm{B_s}} / f_{\mathrm{B}} &=& 1.13(1)\pme{1}{5} \\
f_{\mathrm{D_s}} / f_{\mathrm{D}} &=& 1.11(1)\pme{1}{3} \\
f_\mathrm{K} & = & 150(3)\pme{12}{\ 8}\ \mathrm{MeV}
\end{array}$
&
\renewcommand{\arraystretch}{1.4}
$\begin{array}{rcl}
f_{\mathrm{B^*}}& = & 28(1)^{+3}_{-4}\\
f_{\mathrm{D^*}} &= &8.6(3)^{+5}_{-9} \\
f_{\mathrm{B_s^*}}& = & 25(1)^{+2}_{-3}\\
f_{\mathrm{D_s^*}} &= &8.3(2)\pme{5}{6} \\[2ex]
f_{\mathrm{B^*}} / f_\mathrm{B^*_s} &=& 1.10(2)\pme{2}{6} \\
f_{\mathrm{D^*}} / f_\mathrm{D^*_s} &=& 1.04(1)\pme{2}{4} \\
&&
\end{array}$\\
\end{tabular}
\end{center}
The first error quoted is statistical, the second systematic.
Systematic errors are discussed in Section~\ref{sec:decay_const}, with
the main contributions itemized in Table~\ref{tab:hl_decay_const}.

\section{Details of the calculation}
\label{sec:details}

\subsection{Improved action and operators}
\label{sec:imp-action-ops}

In the Wilson formulation of lattice QCD, the fermionic part of the
action has lattice artifacts of ${\mathcal{O}}(a)$ (where $a$ is the
lattice spacing), while the gauge action differs from the continuum
Yang-Mills action by terms of ${\mathcal{O}}(a^{2})$.  To leading
order in $a$ the Symanzik improvement program for on-shell quantities
involves adding the SW term to the fermionic Wilson action,
\begin{equation}
S_{\mathrm{SW}}=S_{\mathrm{W}} - c_{\mathrm{SW}} \frac{i\kappa}{2}\sum_{x}
{\bar \psi}(x) i \sigma_{\mu\nu} F_{\mu\nu}(x) \psi(x) 
\end{equation}
Full ${\mathcal{O}}(a)$ improvement of on-shell matrix elements also
requires that the currents are suitably improved. The improved vector
and axial currents are
\begin{eqnarray}
\label{eqn:imp_current}
V_{\mu}^\mathrm{I}(x) & = & V_{\mu}(x)+ac_V{\tilde
                         \partial}_{\nu}T_{\mu\nu}(x) \nonumber \\
A_{\mu}^\mathrm{I}(x) & = & A_{\mu}(x) +
                         ac_A{\tilde \partial}_{\mu}P(x)
\end{eqnarray}
where
\begin{eqnarray}
V_{\mu}(x) & = & {\bar \psi}(x) \gamma_{\mu} \psi(x) \nonumber \\
A_{\mu}(x) & = & {\bar \psi}(x) \gamma_{\mu}\gamma_{5} \psi(x)\nonumber \\
P(x)       & = & {\bar \psi}(x) \gamma_{5} \psi(x) \nonumber \\
T_{\mu\nu}(x) & = & {\bar \psi}(x)i\sigma_{\mu\nu} \psi(x) \nonumber
\end{eqnarray}
and ${\tilde \partial_{\mu}}$ is the symmetric lattice derivative.
The generic current renormalisation is as follows $(J=A,V)$:
\begin{equation}
\label{eqn:renorm_current}
J^\mathrm{R} = Z_J(1+b_Jam_q)J^\mathrm{I}
\end{equation}
where $Z_J$ is calculated in a mass-independent renormalisation
scheme.

The bare quark mass, $am_q$, is
\begin{equation}
\label{eqn:amq}
am_q= \frac{1}{2} \left( \frac{1}{\kappa} -
   \frac{1}{\kappa_\mathrm{crit}} \right)
\end{equation}
where $\kappa$ is the hopping parameter.  For non-degenerate currents,
an effective quark mass is used in the definition of the renormalised
current, corresponding to
\begin{equation}
\label{eqn:keff}
\frac{1}{\kappa_\mathrm{eff}}=\frac{1}{2}\left(\frac{1}{\kappa_{1}}+\frac{1}{\kappa_{2}}\right)
\end{equation}
In this renormalisation scheme, the improved quark mass, as used in
chiral extrapolations, is defined as
\begin{equation}
\label{eqn:mimp}
{\widetilde m}_q = m_q (1+b_m am_q)
\end{equation}

\subsection{The static limit} 
The static quark propagator is calculated using the method of Eichten
\cite{eichten}, keeping only the leading term in the expansion of the
propagator in inverse powers of the quark mass:
\begin{equation}
  S_Q(\vec{x},t;\vec{0},0)=\left(
        \theta(t)e^{-m_Qt}\frac{1+\gamma^4}{2} + 
        \theta(-t)e^{m_Qt}\frac{1-\gamma^4}{2} \right)
        \delta^{(3)}(\vec{x})\cl{T}_0(t,0)
\end{equation}
where $\cl{T}_0(t,0)$ is the product of temporal links from $(\vec{0},t)$ to
the origin.
\begin{equation}
  \cl{T}_0(t,0)=\prod_{\tau=1}^{t}U_4^{\dag}(\vec{0},t-\tau)
\end{equation}
The prescription for renormalizing and improving the axial
static-light current is largely analogous to the propagating heavy-light
current~\cite{kurth} case. The difference is that the temporal static axial
current requires covariant derivatives:
\begin{equation}
  (A^\mathrm{stat}_\mathrm{I})_0=A^\mathrm{stat}_0
        +ac_A^\mathrm{stat}\psb_l\gamma_k\gamma_5\frac{1}{2} \left(
        \stackrel{\gets}{D}_k + \stackrel{\gets}{D}_k^{\ast} 
        \right ) \psi_Q
\end{equation}
so that
\begin{equation}
  (A^\mathrm{stat}_\mathrm{R})_0=Z^\mathrm{stat}_A
        (1+b_A^\mathrm{stat} am_\mathrm{q}) (A^\mathrm{stat}_\mathrm{I})_0
\end{equation}
The $c_{\mathrm{A}}^{\mathrm{stat}}$ term is not implemented in this
calculation, but its value is negative which will be significant when
the static point is compared to the infinite mass extrapolation.

\subsection{Definitions of mesonic decay constants}

The pseudoscalar and vector meson decay constants, $f_P$ and
$f_V$, are defined by
\begin{eqnarray}
\langle0|A_{\mu}^\mathrm{R}(0)|P\rangle &=& 
    i p_{\mu} f_P,
\label{eqn:fpdef}\\
\langle0|V_{\mu}^\mathrm{R}(0)|V \rangle &=&
  i \epsilon_{\mu}\frac{M_V^2}{f_V}
\label{eqn:fvdef}
\end{eqnarray}
where $|P\rangle$ is a pseudoscalar meson state with
momentum $p_{\mu}$, while $|V\rangle$ is a vector meson,
with mass $aM_V$ and polarisation vector $\epsilon_\mu$.
$A_\mu^\mathrm{R}$ ($V_\mu^\mathrm{R}$) denotes the
renormalised axial (vector) current, here taken to be the
renormalised, improved local lattice axial (vector) current, defined
via equation~(\ref{eqn:renorm_current}).

\subsection{Simulation details}

In this study, gauge field configurations are generated using a
combination of the over-relaxed~\cite{creutz_or,brown_woch_or} and the
Cabibbo-Marinari~\cite{cabibbo_marinari} algorithms with periodic
boundary conditions at two values of the gauge coupling
$\beta=6/g_0^2$.  At each $\beta$, heavy quark propagators are
computed at four values of the hopping parameter, corresponding to
quarks with masses in the region of the charm quark mass.  For light
quark propagators, three values of $\kappa$ are used, corresponding to
masses around that of the strange quark.  Table~\ref{tab:lattices}
lists the input and derived parameters. Errors quoted in this and
other tables are statistical only unless otherwise specified.
\begin{table}
\begin{center}
\caption{Input and derived parameters. The lattice spacing is set by $f_\pi$.
        \smallskip}
\label{tab:lattices}
\begin{tabular}{lcc}
\hline\hline
                            &  $\beta=6.2$        & $\beta=6.0$        \\
\hline
Volume       & $24^{3}\times 48 $& $16^{3}\times 48 $\\   
$c_{\mathrm{SW}}    $       & 1.614       &     1.769                 \\   
$N_{\mathrm{configs}}$      & 216        &     305                  \\
${a^{-1}} 
\left({\mathrm{GeV}}  
\right)  $                  & $2.66^{+7}_{-7}$  &    $1.91\pme{6}{6}$
\\  
Heavy $\kappa$ & $0.1200,0.1233,0.1266,0.1299$ &
                            $0.1123,0.1173,0.1223,0.1273$\\
Light $\kappa$ & $0.1346,0.1351,0.1353$ & $0.13344,0.13417,0.13455$
\\
$\kappa_{\mathrm{crit}}$    & $0.13581^{+2}_{-1}$ & $0.13525^{+2}_{-1}$\\   
$\kappa_{\mathrm{n}}$    & $0.13577^{+2}_{-1}$ & $0.13519^{+2}_{-1}$\\   
$\kappa_{\mathrm{s}}$ & $0.13479^{+6}_{-6}$ & $0.1338^{+1}_{-1}$\\    
\hline\hline                                    
\end{tabular}
\end{center}
\end{table}
The value of the hopping parameter corresponding to zero quark mass,
$\kappa_\mathrm{crit}$, is taken from ~\cite{QLHS}. The
strange and normal quark masses are fixed using the pseudoscalar meson
masses for the pion and kaon where the lattice spacing has been fixed
from the pion decay constant, $f_\pi$. Statistical errors are
estimated using the bootstrap~\cite{efron} with 1000 re-samplings.

\subsection{Improvement coefficients}

The improvement program requires values for the current and mass
improvement and renormalisation coefficients defined in
equations~(\ref{eqn:imp_current}), (\ref{eqn:renorm_current})
and~(\ref{eqn:mimp}) in Section~\ref{sec:imp-action-ops}. One would
like to use non-perturbative determinations of the coefficients in
order to remove all $\order{a}$ errors. However, different
non-perturbative determinations may give mixing coefficients differing
by terms of $\order{a}$ and normalisation coefficients differing at
$\order{a^2}$. Using a consistently determined set of coefficients
(that is, applying the same improvement condition at each value of the
coupling and a consistent set of improvement conditions for all
coefficients) should enable a smooth continuum extrapolation.

Alternatively, perturbation theory may be used~\cite{alpha_np4_bA},
although this leaves residual discretisation errors of
$\order{\alpha_{\mathrm{s}}^2a}$. Lattice perturbation theory is
improved by using a boosted coupling
$g^2=g_0^2/u^4_0$~\cite{lepage_mack}.  The mean link, $u_0$, is
related to the plaquette expectation value,
$u_0^4=\langle\mathrm{Re\,Tr}\, U_{\mathrm P}\rangle/3$.

Non-perturbative determinations of the improvement coefficients are
available from two groups.  The ALPHA collaboration have determined
the value of $c_{\mathrm{SW}}$~\cite{alpha_np,alpha_np2} using chiral
symmetry and Ward identities in the Schr\"{o}dinger Functional (SF)
formalism.  They have determined $c_A$~\cite{alpha_np2} and $Z_A$,
$Z_V$ and $b_V$~\cite{alpha_np3_Z} in the same scheme and have a
preliminary determination of $c_V$~\cite{sommer_cv,alpha_np5_cv}.
\Bhatt~\cite{bhatta_plb,bhatta_99,Bhatta_2000} have determined all the
improvement coefficients needed to improve and renormalise quark
bilinears, also using Ward identities, but on a periodic lattice with
standard sources. They also use the ALPHA value of $c_{\mathrm{SW}}$
to improve the action.
\begin{table}
\caption{Improvement coefficients from various determinations.
        LANL refers to \Bhatt and BPT is boosted perturbation
        theory.\smallskip}
\label{tab:impcoeff}
\hbox to\hsize{\hss
\begin{tabular}{cccccccc}
\hline\hline
&\multicolumn{3}{c}{$\beta=6.2$} && \multicolumn{3}{c}{$\beta=6.0$}\\
& LANL & ALPHA & BPT && LANL & ALPHA & BPT \\\cline{1-4}\cline{6-8}
$Z_A$ & $0.818(2)(5)$ & $0.807(8)(2)$ & $0.8163$ &&
    $0.807(2)(8)$ & $0.7906(94)$ & $0.8038$\\
$b_A$ & $1.32(3)(4)$ & $-$ & $1.24$&&    
    $1.28(3)(4)$ & $-$ & $1.26$\\
$c_A$ & $-0.032(3)(6)$ & $-0.038(4)$ & $-0.012$ &&
    $-0.037(4)(8)$ & $-0.083(5)$ & $-0.013$ \\ \cline{1-4}\cline{6-8}
$Z_V$ & $0.7874(4)$ & $0.7922(4)(9)$ & $0.7959$ &&
    $0.770(1)$ & $0.7809(6)$ & $0.7820$ \\
$b_V$ &  $1.42(1)(1)$ & $1.41(2)$ & $1.24$ &&
    $1.52(1)$ & $1.54(2)$ & $1.26$ \\
$c_V$ & $-0.09(2)(1)$ & $-0.21(7)$ & $-0.026$ &&
    $-0.107(17)(4)$ & $-0.32(7)$ & $-0.028$ \\
\hline\hline
\end{tabular}\hss}
\end{table}

\begin{figure}
\begin{center}
\epsfig{file=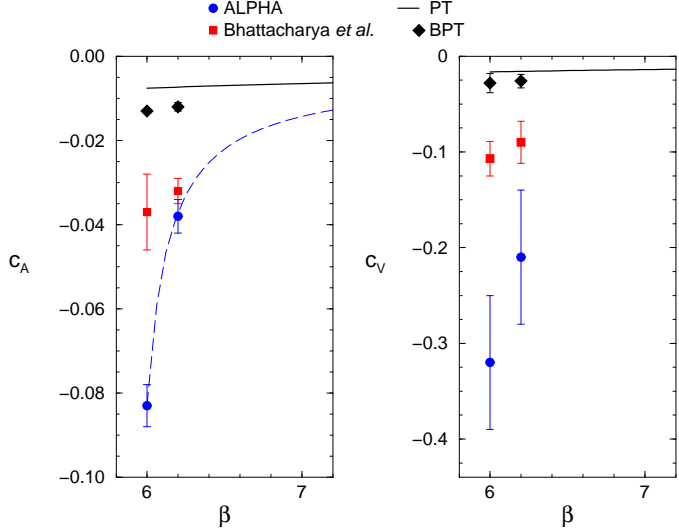,width=0.63\hsize} 
\caption{The improvement coefficients $c_A$ (left) and $c_V$
        (right) as functions of the coupling $\beta$. Note that the vertical 
        scales in the two plots are different.
        The values of the coefficients are shown for the two values of
        the coupling used in this work. The dashed line through the ALPHA
        $c_A$ points is their interpolating function.}
\label{fig:coeff}
\end{center}
\end{figure}

\begin{figure}
\begin{center}
\epsfig{file=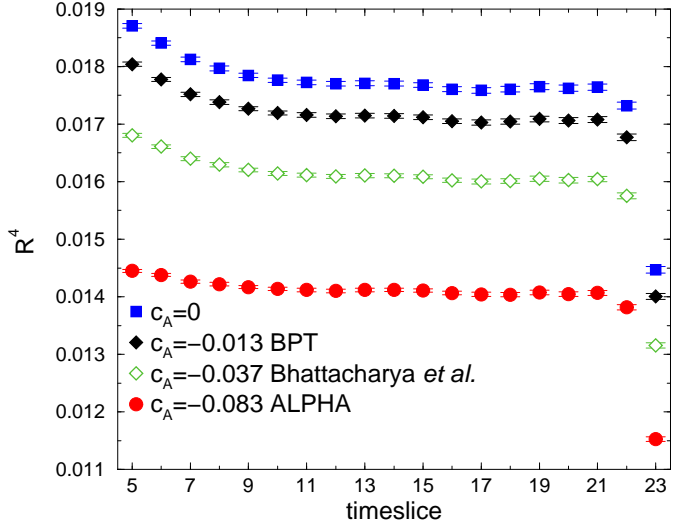,width=0.63\hsize} 
\caption{The mixing of the axial current with the pseudoscalar density.
        $\beta=6.0$, $\kappa_H=0.1123$, $\kappa_L=0.13344$. 
        The ratio plotted is defined in equation (\ref{eqn:fps_tanh}).}
\label{fig:PS_mix}
\end{center}
\end{figure}

The two non-perturbative determinations of the improvement
coefficients give very similar values for the renormalisation
coefficients ($Z$'s) and the quark mass constant $b_V$ (Table
\ref{tab:impcoeff}). However, the mixing coefficients $c_A$ and
$c_V$ differ greatly for the two values of the coupling used
here. This is shown in Figure~\ref{fig:coeff}.  Changes in $c_A$ and
$c_V$ can have a particularly large effect on the extracted values of
the decay constants when the pseudoscalar and vector meson masses are
not small (in lattice units), since the improved current matrix
elements are given by,
\begin{eqnarray}
  \langle 0 | A_4^\mathrm{I} | P \rangle \ &=& \langle 0 | A_4 | P \rangle
        \ + c_A\sinh\left(a M_P\right)
        \langle 0 | P | P \rangle \nonumber\\
  \langle 0 | V_i^\mathrm{I} | V, \epsilon \rangle & = &
        \langle 0 | V_i | V, \epsilon\rangle +
        c_V\sinh \left(a M_V\right )
         \langle 0 | T_{i4} | V, \epsilon\rangle
\end{eqnarray}
when the ground state is isolated.  In particular at $\beta=6.0$ the
heavy-light meson mass at the heaviest kappa is somewhat bigger than
one such that $\sinh aM \sim 1.3$.  This is illustrated for the
pseudoscalar decay constant in Figure~\ref{fig:PS_mix}. The figure
shows the ratio $R^4$ (between PA and PP correlators, defined in
equation~(\ref{eqn:fps_tanh}) and proportional to $f_P$), for several
different values of $c_A$ at $\beta=6.0$. It can be seen that using
the NP value of $c_A$ from the ALPHA collaboration decreases $R^4$ by
$\sim 20\%$ relative to the case $c_A = 0$. The plot also shows the
ratio determined using the NP value of $c_A$ from \Bhatt

At $\beta=6.0$ the disagreement between ALPHA and \Bhatt in the mixing
coefficients is striking.  \Bhatt try to estimate some of the cut-off
effects in their determination of $c_A$ by looking at the difference
between two- and three-point derivatives. They also note that since
$a^2$ only halves between $\beta=6.0$ and $\beta=6.2$, where
disagreement is substantially decreased, even higher order cut-off
effects are playing a dominant r\^{o}le. Collins and
Davies~\cite{Glasgow_cA} have examined the effect of using
higher-order derivatives in determining $c_A$ on the data set used in
this work and point out that the difference in the value of $c_A$ from
different orders of derivative disappears once the chiral limit is
taken.  There is clearly a large $\order{a}$ ambiguity in the value of
$c_A$.

\Bhatt have computed all the coefficients needed for this calculation
in a consistent fashion and their determinations are used here for
that reason. The values of the mixing coefficients $c_A$ and $c_V$ are
an order of magnitude larger in the ALPHA determinations than in BPT
at $\beta=6.0$. This has a large impact on the HQS relation between,
and scaling of, the decay constants. When the values of the mixing
coefficients from \Bhatt are used the HQS relation is satisfied and
there is good scaling.

The ALPHA collaboration compute $b_V$ in the SF scheme at very small
quark masses and in \Bhatt the heaviest quark mass is
$am_q\sim0.13$. One may question whether these coefficients can be
used around the charm mass: at $\beta=6.0$, $am_c\sim0.75$ and at
$\beta=6.2$, $am_c\sim0.5$. However, the effective normalisation of
the vector~\cite{B2pi_plb,chrism_2000} and axial~\cite{FF_prep}
currents has been measured at these quark masses and found to be in
good agreement with the determinations of $Z_V$ and
$b_V$\footnote{Indeed \Bhatt compare their results at $\beta=6.2$ to
\cite{B2pi_plb}}.

A value for $b_m$ is required to compute the rescaled quark mass used
in the chiral extrapolations.  Whilst there is a non-perturbative
determination \cite{bmnonpert}, it is only at one value of the
coupling, and so this work follows \cite{QLHS} and uses the BPT value
with which the non-perturbative value is consistent.

Expressions for the static improvement and renormalisation coefficients
$c^\mathrm{stat}_A$, $b^\mathrm{stat}_A$ and $Z^\mathrm{stat}_A$, have been
derived in perturbation theory~\cite{kurth},
\begin{eqnarray}
\label{eqn:stat_imp_coeffs}
Z_A^\mathrm{stat}&=&1.0+\left( \frac{\ln(a\mu)}{4\pi^2} - 0.137(1) \right) g^2
        +\order{g^4} \\\nonumber
b_A^\mathrm{stat} &=&\frac{1}{2} -0.056(7)g_0^2 + \order{g_0^4} \\\nonumber
c_A^\mathrm{stat} & =& -\frac{1}{4\pi}\times 1.01(5)g_0^2
\end{eqnarray}
where $g_0$ is the bare lattice coupling.  The one-loop PT and NP
renormalisation schemes have been matched at the scale
$\mu=m_b$~\cite{kurth_priv,sommer_review} in the
$\overline\mathrm{MS}$ scheme. The coupling $g$ is then evaluated in
the $\overline\mathrm{MS}$ scheme~\cite{pdg2000} at the scale
$m_b$. The values for $b^\mathrm{stat}_A$ and $c^\mathrm{stat}_A$ have
been evaluated with the boosted coupling at $\beta=6.0$. The values
are
\begin{eqnarray}
  Z_A^\mathrm{stat}&=&0.663\pme{17}{15} \\\nonumber
  b_A^\mathrm{stat}&=&0.4057 \\\nonumber
  c_A^\mathrm{stat}&=&-0.1354
\end{eqnarray}
where the errors on $Z_A^\mathrm{stat}$ reflect the uncertainty in 
$m_b=4.0-4.4$ GeV~\cite{pdg2000} and in the $\overline\mathrm{MS}$ coupling.

\subsection{Extraction of masses and decay constants of heavy-light mesons}

The pseudoscalar meson masses are extracted from the asymptotic
behaviour of the two-point correlation functions,
\begin{eqnarray}
  C_{PP}^{SS}(t,\vec{p}) &=& 
  \sum_{\vec{x}} e^{-i\vec{p}{\cdot}\vec{x}}
  \langle \Omega_P^S(t,\vec{x}) 
  \Omega_P^{S\dagger}(0,\vec{0}) \rangle \\\nonumber
  &\stackrel{t\to\infty}\to&
  {\left(Z^S_P(\vec{p})\right)^2 \over 2aE_P} 
  \cosh(aE_P (t-T/2)) e^{-aE_P T/2}
\end{eqnarray}
where $T=48$ and $aE_P$ is the energy of the lowest lying meson
destroyed by the operator $\Omega^S_P$ and created by
$\Omega_P^{S\dagger}$. The superscript $S$ denotes a smeared, or
spatially extended, interpolating field operator constructed using the
gauge invariant technique described in~\cite{boyling_p}.
$Z_P^S(\vec{p})$ is the overlap of the operator with the pseudoscalar
state given by $Z_P^S(\vec{p})$ = $\langle 0 |\Omega_P^S(0,\vec{0})|
{\mathrm P}(\vec{p}) \rangle$.  Fit ranges are established by
inspection of the time-dependent effective mass.
Table~\ref{tab:hlmass} shows the best fit masses in each case. A
similar procedure is used to extract the masses of vector mesons from
correlation functions constructed with the appropriate vector
operators.
\begin{table}
\begin{center}
\caption{Pseudoscalar and vector masses in lattice units at
$\beta=6.2$ and $\beta=6.0$. Fit ranges are $12-22$ at $\beta=6.2$ and
$10-22$ (P), $10-21$ (V) at $\beta=6.0$.\smallskip}
\label{tab:hlmass}
\begin{tabular}{cccc}\hline\hline
\multicolumn{4}{c}{$\beta=6.2$} \\
$\kappa_\mathrm{H}$ & $\kappa_\mathrm{L}$ & $aM_P$ & $aM_V$\\\hline
     & $0.1346$ &$0.841^{+1}_{-1}$ & $0.871^{+2}_{-2}$     \\
$0.1200$ & $0.1351$&$0.823^{+2}_{-1}$&$0.856^{+2}_{-2}$ \\
    & $0.1353$&$0.817^{+2}_{-1}$&  $0.848^{+3}_{-2}$   \\
\hline
     & $0.1346$&$0.739^{+1}_{-1}$ & $0.775^{+2}_{-2}$ \\
$0.1233$ & $0.1351$&$0.721^{+2}_{-1}$& $0.759^{+2}_{-2}$ \\
     & $0.1353$&$0.714^{+2}_{-1}$&  $0.752^{+3}_{-2}$ \\
\hline
     & $0.1346$&$0.628^{+1}_{-1}$& $0.673^{+2}_{-2}$ \\
$0.1266$ & $0.1351$&$0.609^{+1}_{-1}$& $0.656^{+2}_{-2}$ \\
     & $0.1353$&$0.602^{+2}_{-1}$& $0.650^{+3}_{-2}$ \\
\hline
     & $0.1346$&$0.505^{+1}_{-1}$& $0.563^{+2}_{-2}$ \\
$0.1299$ & $0.1351$&$0.484^{+1}_{-1}$& $0.546^{+2}_{-2}$ \\
     & $0.1353$&$0.476^{+1}_{-1}$& $0.540^{+2}_{-2}$ \\\hline\hline
\end{tabular}
\hfill
\begin{tabular}{cccc}\hline\hline
\multicolumn{4}{c}{$\beta=6.0$} \\
$\kappa_\mathrm{H}$ & $\kappa_\mathrm{L}$ & $aM_P$ & $aM_V$\\\hline
     & $0.13344$ &$1.145^{+2}_{-1}$ & $1.188^{+2}_{-2}$     \\
$0.1123$ & $0.13417$&$1.121^{+2}_{-2}$&$1.166^{+3}_{-3}$ \\
    & $0.13455$&$1.110^{+3}_{-2}$&  $1.158^{+4}_{-4}$   \\
\hline
     & $0.13344$&$1.006^{+2}_{-1}$ & $1.056^{+2}_{-2}$ \\
$0.1173$ & $0.13417$&$0.981^{+2}_{-2}$& $1.034^{+3}_{-2}$ \\
     & $0.13455$&$0.969^{+2}_{-2}$&  $1.026^{+4}_{-4}$ \\
\hline
     & $0.13344$&$0.851^{+1}_{-1}$& $0.915^{+2}_{-2}$ \\
$0.1223$& $0.13417$&$0.825^{+2}_{-1}$& $0.892^{+3}_{-2}$ \\
     & $0.13455$&$0.811^{+2}_{-2}$& $0.883^{+4}_{-3}$ \\
\hline
     & $0.13344$&$0.675^{+1}_{-1}$& $0.759^{+2}_{-2}$ \\
$0.1273$ & $0.13417$&$0.646^{+2}_{-1}$& $0.736^{+3}_{-2}$ \\
     & $0.13455$&$0.631^{+2}_{-1}$& $0.727^{+4}_{-3}$ \\\hline\hline
\end{tabular}
\end{center}
\end{table}

The decay constants are extracted from the large time behaviour of
different two-point correlation functions at zero momentum. The PA
correlation function is used for the pseudoscalar decay constant:
\begin{eqnarray}
C_{PA}^{SL}(t,\vec{0}) &=& 
  \sum_{\vec{x}} 
\langle A_4^\mathrm{I}(t,\vec{x})
  \Omega_P^{S\dagger}(0,\vec{0}) \rangle \\ \nonumber
 &\stackrel{t\to\infty}\to&
  \frac{Z_A^L  Z_P^S}{2aM_P}
{\sinh(aM_P (t-T/2)) e^{-aM_P  T/2} } 
\end{eqnarray}
where $A_4^\mathrm{I}$ is the time component of the improved axial
current operator defined in equation~(\ref{eqn:imp_current}).  The
superscript $L$ on the correlator denotes a local operator, in this
case the axial current. $Z^L_A =
\langle0|A_4^\mathrm{I}|P(\vec0)\rangle$ is the overlap of the local
axial operator with the pseudoscalar state, from which the decay
constant is extracted using equations~(\ref{eqn:renorm_current})
and~(\ref{eqn:fpdef}).

Extraction of the vector decay constant involves the large 
time behaviour of the VV correlation function:
\begin{eqnarray}
C_{VV}^{SL}(t,\vec{0}) &=& 
\sum_j\sum_{\vec{x}} 
\langle V_j^\mathrm{I}(t,\vec{x}) 
\Omega_{V_j}^{S\dagger}(0,\vec{0}) \rangle \\\nonumber
 &\stackrel{t\to\infty}\to&
 \frac{Z_V^LZ_V^S}{2aM_V}
 \cosh(aM_V (t-T/2)) e^{-aM_V T/2} 
\end{eqnarray}
where $V_j^\mathrm{I}$ is a spatial component of the improved local vector
current operator. Again, $S$ denotes a smeared or spatially extended
interpolating field operator and $L$ a local operator. The factor $Z_V^L$
is the overlap of the local vector operator with the vector state,
$Z^L_V =\sum_r \epsilon^r_k
\langle0|V_k^\mathrm{I}|V(\vec0,\vec\epsilon\,)\rangle$, from which the vector
decay constant can be extracted via equation (\ref{eqn:fvdef}).

Matrix elements proportional to the decay constants can be extracted
from ratios of correlation functions. The ratio
\begin{equation}
\label{eqn:fps_tanh}
R^4\equiv\frac{C^{SL}_{PA}(t)}{C^{SS}_{PP}(t)}
  \stackrel{t\to\infty}\to
  \frac{Z_A^L}{Z_P^S} \tanh (aM_P(T/2-t))
\end{equation}
is used for the pseudoscalar case and is shown in
Figure~\ref{fig:PS_mix} for various values of $c_A$.  The
vector decay constant is determined using
\begin{equation}
\label{eqn:fV_ratio}
\frac{C^{SL}_{VV}(t)}{C^{SS}_{VV}(t)}\stackrel{t\to\infty}\to
  \frac{Z_V^L}{Z_V^S}
\end{equation}
Results for the decay constants are shown in Table~\ref{tab:hldecon}.

An alternative method is to fit several correlators simultaneously
allowing an estimation of the contamination of the ground state signal
by excited states.  An eight-parameter fit is made to $C^{SS}_{PP}$,
$C^{SL}_{PA}$ and $C^{SL}_{PP}$, allowing for ground and first excited
state contributions. Results for the pseudoscalar decay constant are
consistent with the single ratio fit. The ratio method is used for
central values in the following but the difference from the
multi-exponential fits is quoted in Table~\ref{tab:sys_error} below as
one measure of systematic error.
\begin{table}
\begin{center}
\caption{Pseudoscalar and vector decay constants in lattice units. At 
        $\beta=6.2$ the fit ranges are $14-21$ (P) and $15-23$ (V).
        At $\beta=6.0$ the fit ranges are $14-21$ (P) and $16-23$ (V).
        \smallskip}
\label{tab:hldecon}
\begin{tabular}{cccc}\hline\hline
\multicolumn{4}{c}{$\beta=6.2$}\\
$\kappa_\mathrm{H}$ & $\kappa_\mathrm{L}$ & $af_P$ &  $f_V$   \\\hline
     & $0.1346$ &$0.0892^{+8}_{-8}$ & $9.03^{+10}_{-\ 9}$     \\
$0.1200$ & $0.1351$&$0.0847^{+8}_{-9}$&$9.28^{+12}_{-12}$   \\
    & $0.1353$&$0.0832^{+9}_{-9}$&  $9.37^{+14}_{-13}$     \\
\hline
     & $0.1346$&$0.0873^{+7}_{-8}$ & $7.98^{+\ 8}_{-\ 8}$    \\
$0.1233$ & $0.1351$&$0.0829^{+7}_{-8}$& $8.17^{+10}_{-\ 9}$   \\
     & $0.1353$&$0.0814^{+8}_{-9}$&  $8.23^{+11}_{-10}$  \\
\hline
     & $0.1346$&$0.0847^{+6}_{-7}$& $6.89^{+\ 6}_{-\ 6}$  \\
$0.1266$& $0.1351$&$0.0805^{+6}_{-8}$& $7.00^{+\ 8}_{-\ 7}$  \\
     & $0.1353$&$0.0790^{+7}_{-9}$& $7.03^{+\ 9}_{-\ 8}$ \\
\hline
     & $0.1346$&$0.0804^{+6}_{-7}$& $5.76^{+\ 5}_{-\ 5}$  \\
$0.1299$ & $0.1351$&$0.0765^{+6}_{-8}$& $5.79^{+\ 6}_{-\ 5}$  \\
     & $0.1353$&$0.0751^{+6}_{-9}$& $5.79^{+\ 7}_{-\ 6}$ \\
\hline\hline
\end{tabular}
\hfill
\begin{tabular}{cccc}\hline\hline
\multicolumn{4}{c}{$\beta=6.0$}\\
$\kappa_\mathrm{H}$ & $\kappa_\mathrm{L}$ & $af_P$ & $f_V$ \\\hline
     & $0.13344$ &$0.1244^{+10}_{-11}$ & $8.79^{+11}_{-\ 9}$    \\
$0.1123$ & $0.13417$&$0.1187^{+12}_{-13}$&$8.94^{+14}_{-12}$  \\
    & $0.13455$&$0.1167^{+15}_{-16}$&  $8.89^{+19}_{-14}$    \\
\hline
     & $0.13344$&$0.1217^{+10}_{-10}$ & $7.80^{+\ 9}_{-\ 8}$  \\
$0.1173$ & $0.13417$&$0.1162^{+10}_{-11}$& $7.89^{+11}_{-10}$  \\
     & $0.13455$&$0.1142^{+12}_{-14}$&  $7.84^{+16}_{-11}$ \\
\hline
     & $0.13344$&$0.1170^{+\ 7}_{-\ 9}$& $6.74^{+\ 7}_{-\ 6}$  \\
$0.1223$& $0.13417$&$0.1118^{+\ 8}_{-10}$& $6.79^{+\ 9}_{-\ 8}$  \\
     & $0.13455$&$0.1097^{+10}_{-12}$& $6.72^{+12}_{-\ 9}$  \\
\hline
     & $0.13344$&$0.1102^{+\ 6}_{-\ 9}$& $5.62^{+\ 6}_{-\ 5}$  \\
$0.1273$ & $0.13417$&$0.1053^{+\ 7}_{-\ 9}$& $5.59^{+\ 7}_{-\ 6}$  \\
     & $0.13455$&$0.1032^{+\ 8}_{-11}$& $5.51^{+\ 9}_{-\ 7}$ \\
\hline\hline
\end{tabular}
\end{center}
\end{table}

\subsubsection{The static-light axial current}
The static-light axial current can be extracted from axial-axial correlation
functions. The local-local and local-smeared correlation functions were
used in this work, with large-time behaviour given by 
\begin{eqnarray}
C_{AA}^{LL}{}(t,\vec{0}) &\ =\ 
  \sum_{\vec{x}} 
\left \langle A^L_4(t,\vec{x})
  A^{L\dagger}_4(0,\vec{0}) \right \rangle 
 &\ \stackrel{t\to\infty}\to \   (Z_A^L)^2e^{-a\Delta E t} \\ \nonumber
C_{AA}^{LS}(t,\vec{0}) &\ =\ 
  \sum_{\vec{x}} 
\left \langle A^S_4(t,\vec{x})
  A^{L\dagger}_4(0,\vec{0}) \right \rangle 
 &\ \stackrel{t\to\infty}\to \   Z^S_AZ_A^Le^{-a\Delta E t} \\ \nonumber
\end{eqnarray}
where $\Delta E$ is the unphysical difference between the mass of the
meson and the mass of the bare heavy quark. The static-light
amplitude, $Z_L$, is 
\begin{equation}
  Z_L=af_P^\mathrm{stat} \sqrt{\frac{aM_P^\mathrm{stat}}{2}}
\end{equation}
The smearing function is again the one described in \cite{boyling_p}.

The static-light correlation functions were generated at only one
value of the coupling, $\beta=6.0$, and without the covariant
derivative operators necessary to improve the current. 
The assumption that the matrix element of the improvement term is of
the same order of magnitude as the primary term leads to
$\order{10\%}$ error in the static point associated with the absence
of improvement.

\begin{figure}
\begin{center}
\epsfig{file=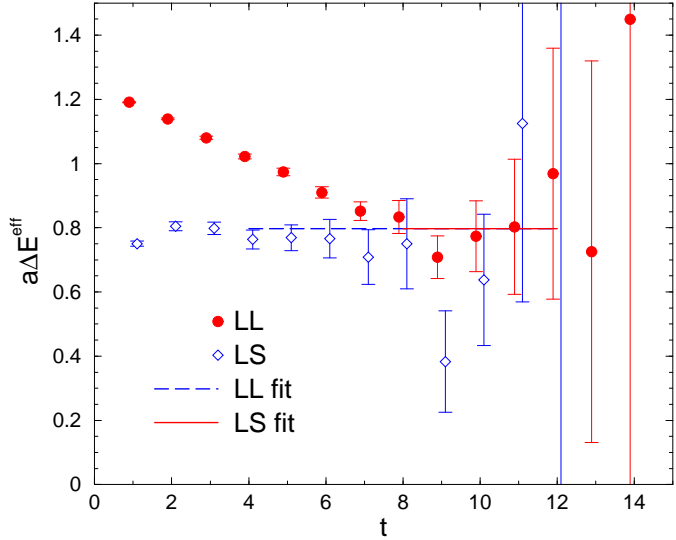,width=0.63\hsize}
\caption{The effective mass plot for the LL and LS static-light axial
        correlation function. $\kappa_L=0.13344$. The lines on the
        plot show the value of the fit, and the fit ranges.}
\label{fig:stat_effmass}
\end{center}
\end{figure}

The signal of the static-axial current is very quickly overwhelmed by
statistical noise, making the fit difficult. Simultaneously fitting to
both the local-local and local-smeared correlator gives a better
estimate of the local current. This is shown in Figure
\ref{fig:stat_effmass}. The renormalised values of 
$Z_L$ (without the current improvement term) are shown in Table
\ref{tab:fB_stat}.
\begin{table}[!ht]
\begin{center}
\caption{Values of $a\Delta E$ and $Z_L$ from simultaneous fits to the
        LL and LS correlation functions. The fit ranges are 8-12 for the
        LL and 4-12 for the LS.\smallskip}
\label{tab:fB_stat}
\begin{tabular}{ccc}
\hline\hline
 $\kappa_L$ & $Z_L$ & $a\Delta E$ \\\hline
   0.13344 &$0.30(2)$&$0.80(2)$\\
   0.13417 &$0.28(2)$&$0.78(3)$\\
   0.13455 &$0.26(3)$&$0.76(3)$ \\\hline\hline
\end{tabular}
\end{center}
\end{table}

\subsection{Extraction of masses and decay constants of light-light mesons}

The masses and decay constants of the light-light pseudoscalar mesons
are extracted from a simultaneous fit to three correlation functions:
$C^{SS}_{PP}$, $C^{SL}_{PP}$ and $C^{SL}_{PA}$. The smearing function
used for the light-lights is the `fuzzing' of
reference~\cite{fuzzing}. The results for the decay constants are
shown in Table~\ref{tab:light_ps_decay}. The results for the
pseudoscalar masses are compatible with those in \cite{QLHS} where a
similar fit was used.

\begin{table}
\begin{center}
\caption{The light-light pseudoscalar decay constants. The fit ranges
        are $8-22$ at both values of the coupling, except for the
        lightest mass combination at $\beta=6.0$ where the fit range was
         $7-23$.\smallskip}
\label{tab:light_ps_decay}
\begin{tabular}{ccc}
\hline\hline
\multicolumn{3}{c}{$\beta=6.2$}\\
$\kappa_1$ & $\kappa_2$ & $af_P$ \\\hline
$0.13460$& $0.13460$& $0.0640\pme{7}{7}$\\
$0.13510$& $0.13460$&$0.0610\pme{6}{7}$ \\
$0.13530$& $0.13460$&$0.0599\pme{6}{7}$ \\
$0.13510$& $0.13510$&$0.0587\pme{5}{6}$ \\ 
$0.13530$& $0.13510$&$0.0575\pme{6}{6}$ \\
$0.13530$& $0.13530$&$0.0559\pme{6}{7}$ \\\hline\hline
\end{tabular}
\hfill
\begin{tabular}{ccc}
\hline\hline\multicolumn{3}{c}{$\beta=6.0$}\\
 $\kappa_1$ & $\kappa_2$ & $af_P$ \\\hline
 $0.13344$& $0.13344$& $0.0881\pme{10}{10}$\\
 $0.13417$& $0.13344$& $0.0841\pme{10}{\ 9}$\\
 $0.13455$& $0.13344$& $0.0821\pme{10}{10}$\\
$0.13417$& $0.13417$& $0.0798\pme{\ 9}{11}$\\
 $0.13455$& $0.13417$& $0.0775\pme{\ 9}{15}$\\
 $0.13455$& $0.13455$&$0.0730\pme{17}{18}$ \\\hline\hline
\end{tabular}
\end{center}
\end{table}

\section{Extrapolation and interpolation in the quark masses}
\label{sec:extrap}

Extrapolation or interpolation in quark masses must be performed to
extract physical masses and decay constants.  For heavy quarks, the
presence of $\order{a^2m_Q^2}$ lattice artefacts when using the SW
action with the NP improved renormalisation scheme imposes a
constraint $am_Q<1$ on the quark masses that can be studied. This
limits hadron masses to $\sim 2$ GeV or slightly greater at
$\beta=6.0$, see Table~\ref{tab:quark_masses}. Input light quark
masses are kept above $m_\mathrm{s}/2$ to avoid critical slowing down of
quark propagator calculations and the possible appearance of finite
volume effects. Interpolations to $m_\mathrm{c}$ and $m_\mathrm{s}$
are needed, together with extrapolations for $m_\mathrm{b}$ and the
light quarks.

\subsection{The light-light sector} 

The dependence of the light pseudoscalar decay constant on the light 
pseudoscalar meson mass is described by chiral perturbation theory,
giving the following ansatz: 
\begin{equation} 
  af_P = c_0 + c_1 (aM_P)^2 + c_2(aM_P)^4  
\end{equation} 
The value of $af_P$ for which
\begin{equation}
 \frac{af_P}{aM_P}=\left(\frac{f_\pi}{m_\pi}\right)_\mathrm{expt}
\end{equation}
can then be determined and compared to the experimental value of
$f_\pi$, taken from~\cite{pdg2000},
to fix the lattice spacing. At both values of the coupling,
quadratic and linear fits in $(aM_P)^2$ give essentially the same
answer for the lattice spacing. Thus the data satisfy lowest-order
chiral perturbation theory. Because of quenching and other systematic effects,
determinations of the lattice spacing from different quantities
disagree. In this work the lattice spacing is
set by $f_\pi$. Lattice spacings fixed by the Sommer scale, $r_0$,
\cite{sommer_r0,wittig_r0}, and $m_\rho$ as determined by \cite{QLHS}
are used to estimate systematic error from this source. The values are
shown in Table~\ref{tab:latt_a}.

\begin{table}
\begin{center}
\caption{The inverse lattice spacing, $a^{-1}$ (GeV)\smallskip}
\label{tab:latt_a}
\begin{tabular}{ccc}
\hline\hline
 & $\beta=6.2$ & $\beta=6.0$  \\\hline
$f_\pi$ &$2.66\pme{7}{7}$ & $1.91\pme{6}{6}$ \\
$r_0$ &$2.91\pme{1}{1}$ & $2.12\pme{1}{1}$ \\
$m_\rho$ &$2.54\pme{4}{9}$ & $1.89\pme{3}{5}$ \\
\hline\hline
\end{tabular}
\end{center}
\end{table}

The light, or `normal', quark mass $m_{\mathrm n}$, defined by
$m_{\mathrm n} \equiv (m_{\mathrm u} + m_{\mathrm d})/2$, and strange
quark mass $m_{\mathrm s}$ are determined using the lowest-order
chiral perturbation theory relation for the mass of a pseudoscalar
meson with quark content $q_1$ and $q_2$,
\begin{equation}
(a m_{\mathrm{P}})^2 = B (a\widetilde m_{q_1} + a\widetilde m_{q_2}) 
\end{equation}
where the rescaled quark mass $\widetilde{m}_\mathrm{q}$ is defined in
equation (\ref{eqn:mimp}). The value of $\kappa_{\mathrm{crit}}$ is
taken from a previous UKQCD calculation~\cite{QLHS} and is listed in
Table~\ref{tab:lattices}. The value of the hopping parameter corresponding
to the normal quark mass is set by the charged pion according to
\begin{equation}
  a^2(m^2_\pi)_\mathrm{expt} = 2 B a \widetilde m_\mathrm{n}
\end{equation}
and that for the strange quark mass by
\begin{equation}
  a^2(m^2_K)_I = B ( a \widetilde m_\mathrm{n}  + a \widetilde m_\mathrm{s} )
\end{equation}
where
\begin{equation}
  (m^2_K)_I= \frac{1}{2}(m_{K^\pm}^2 + m^2_{K^0} )
\end{equation}
with the lattice spacing set by $f_\pi$. These hopping parameters are
also listed in Table~\ref{tab:lattices}.

\begin{figure}
\begin{center}
\epsfig{file=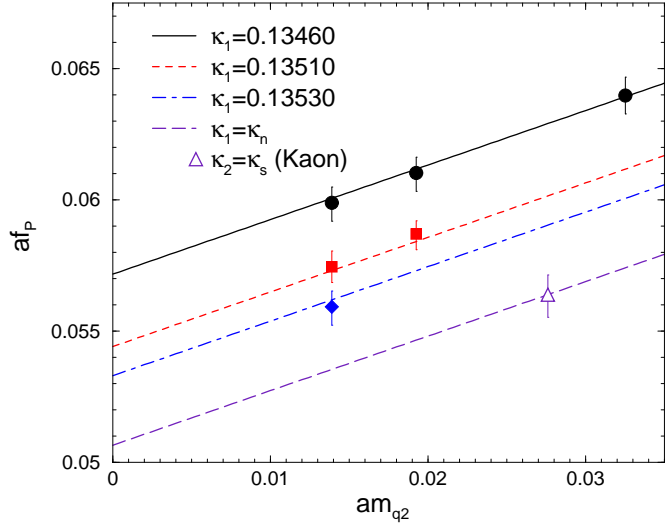,width=0.63\hsize}
\caption{Dependence of the light pseudoscalar decay constant $f_P$ on
         quark masses. The lines give $f_P$ as a function of one of
         the quark masses, $a\tilde m_{q_2}$, for fixed values of the
         other quark mass, $a\tilde m_{q_1}$. Filled symbols are the
         lattice data points. The lowest line shows the extrapolation
         of $a\tilde m_{q_1}$ to the normal quark mass, with the open
         triangle giving $f_\mathrm{K}$ from the position of the kaon
         on this line.}
\label{fig:fK_extrap}
\end{center}
\end{figure}

The ansatz for the dependence of the light pseudoscalar decay constant on
quark masses is:
\begin{equation}
  a f_P = f_0 + f_1 (a\widetilde m_{q_1} + a\widetilde m_{q_2}) 
\end{equation}
Figure~\ref{fig:fK_extrap} shows the results of fitting to this
ansatz and extrapolating to $f_\mathrm{K}$.

\subsection{Chiral extrapolations of heavy-light masses and decay constants}

\begin{figure}
\begin{center}
\epsfig{file=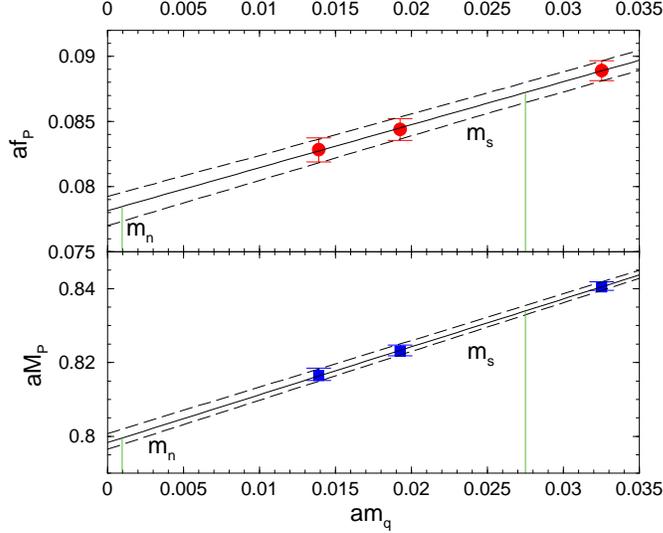,width=0.63\hsize} 
\caption{The chiral extrapolation of the pseudoscalar decay constant (top)
        and pseudoscalar mass (bottom) against rescaled light quark
        mass. $\beta=6.2$ and $\kappa_H=0.1200$. The vertical lines
        show the strange and normal quark masses.}
\label{fig:chiral_extrap}
\end{center}
\end{figure}

A linear dependence of heavy-light masses and decay constants on the
light quark mass is assumed:
\begin{equation}
 a K_{i}=\alpha_i + \beta_i a\widetilde{m}_\mathrm{q}
\end{equation}
where $K_i$ is $f_P$, $f_V/a$, $m_P$ or $m_V$.
Some example extrapolations are shown in
Figure~\ref{fig:chiral_extrap}. The results for the extrapolated
quantities are shown in Tables~\ref{tab:chiral_extrap_B62}~
and~\ref{tab:chiral_extrap_B60}.
\begin{table}
\begin{center}
\caption{Masses and decay constants at physical light quark
masses. $\beta=6.2$.\smallskip}
\label{tab:chiral_extrap_B62}
\begin{tabular}{cccccc}\hline\hline
$\kappa_H$&$\kappa_L$& $aM_P$ & $af_P$ & $aM_V$ & $f_V$\\\hline
 & $\kappa_n$ & $0.800(2)$& $0.079(1)\ $& $0.832(4)$ &$9.6(2) $\\
\raisebox{1.5ex}[0pt]{$0.1200$} & $\kappa_s$
 & $0.828(2)$& $0.0876(8)$& $0.860(2)$& $9.1(1)$ \\\hline
 & $\kappa_n$ & $0.696(2)$& $0.077(1)$& $0.736(3)$& $8.4(1)$ \\
\raisebox{1.5ex}[0pt]{$0.1233$} & $\kappa_s$
 & $0.726(1)$& $0.0857(8)$& $0.764(2)$& $8.05(9)$ \\\hline
 & $\kappa_n$ & $0.583(2)$& $0.075(1)$& $0.634(3)$& $7.1(1)$ \\
\raisebox{1.5ex}[0pt]{$0.1266$} & $\kappa_s$
 & $0.615(1)$& $0.0832(7)$& $0.661(2)$& $6.93(7)$ \\\hline
 & $\kappa_n$ & $0.455(2)$& $0.071(1)$& $0.523(3)$& $5.8(1)$ \\
\raisebox{1.5ex}[0pt]{$0.1299$} & $\kappa_s$
 & $0.491(1)$& $0.0789(7)$& $0.551(1)$& $5.77(6)$ \\\hline\hline
\end{tabular}
\end{center}
\end{table}
\begin{table}
\begin{center}
\caption{Masses and decay constants at physical light quark
masses. $\beta=6.0$.\smallskip}
\label{tab:chiral_extrap_B60}
\begin{tabular}{cccccc}\hline\hline
$\kappa_H$&$\kappa_L$& $aM_P$ & $af_P$ & $aM_V$ & $f_V$\\\hline
 & $\kappa_n$ & $1.087(4)$& $0.112(2)$& $1.138(5)$& $9.0(2)$ \\
\raisebox{1.5ex}[0pt]{$0.1123$} & $\kappa_s$
& $1.125(2)$& $0.122(1)$& $1.171(3)$& $8.8(1)$ \\\hline
 & $\kappa_n$ & $0.945(3)$& $0.109(2)$& $1.005(5)$& $7.9(2)$ \\
\raisebox{1.5ex}[0pt]{$0.1173$} & $\kappa_s$
& $0.985(2)$& $0.119(1)$& $1.039(3)$& $7.8(1)$ \\\hline
 & $\kappa_n$ & $0.786(3)$& $0.105(1)$& $0.862(5)$& $6.8(1)$ \\
\raisebox{1.5ex}[0pt]{$0.1223$} & $\kappa_s$
& $0.829(2)$& $0.115(1)$& $0.897(3)$& $6.75(8)$ \\\hline
 & $\kappa_n$ & $0.603(2)$& $0.099(1)$& $0.704(5)$& $5.5(1)$ \\
\raisebox{1.5ex}[0pt]{$0.1273$} & $\kappa_s$
& $0.651(2)$& $0.1080(9)$& $0.740(3)$& $5.60(7)$ \\\hline\hline
\end{tabular}
\end{center}
\end{table}

\subsection{Heavy quark extrapolations}

Heavy Quark Symmetry (HQS) implies asymptotic scaling
laws~\cite{neubert_fp} for the decay constants in the infinite heavy
quark mass limit. Away from this limit, heavy quark effective theory
ideas motivate the following ans\"atze for the dependence on the heavy
meson masses:
\begin{eqnarray}
\label{eqn:HQS_fp}
\Phi_P(M_P) &\equiv& 
     \Theta(M_\mathrm{B},M_P)f_P\sqrt M_P = 
     \gamma_P \left( 1 + \frac{\delta_P}{M_P} 
     + \frac{\eta_P}{M_P^2} \right)
\\
\label{eqn:HQS_fv}
\Phi_V(M_V) &\equiv& 
  \Theta(M_\mathrm{B},M_V)\frac{M_V}{f_V}\sqrt M_V = 
  \gamma_V \left( 1 + \frac{\delta_V}{M_V} + 
  \frac{\eta_V}{M_V^2} \right)
\end{eqnarray}
where $\Theta$ denotes logarithmic corrections given at
leading order by~\cite{neubert_fpfv},
\begin{equation}
\label{eqn:theta}
  \Theta(M_\mathrm{B},M)=
  \left (\frac{\alpha(M)}{\alpha(M_\mathrm{B})} \right )^{2/\beta_0}
\end{equation}
Here, $\beta_0$ is the one-loop QCD beta function coefficient, equal
to $11$ in the quenched approximation, and
$\Lambda^{(4)}_{\overline{\mathrm{MS}}}=295$ MeV~\cite{alpha_s}. 

HQS also relates the pseudoscalar and vector decay constants as
follows~\cite{neubert_fpfv};
\begin{equation}
\label{eqn:HQS_fpfv}
  U(\overline{M})\equiv \frac{f_Vf_P}{\overline{M}}=
        \left(1+\frac{8}{3}\frac{\alpha_s(\overline{M})}{4\pi} 
        + {\mathcal O}(1/\overline{M}) \right)
\end{equation}
where $\overline{M}\equiv (M_P+3M_V)/4$ is the spin-averaged heavy
meson mass.  The one-loop factor $\Theta$ in
equations~(\ref{eqn:HQS_fp}) and~(\ref{eqn:HQS_fv}) cancels in the
ratio in equation~(\ref{eqn:HQS_fpfv}).  Higher-order QCD corrections
produce the term proportional to~$\alpha_\mathrm{s}$.
$\widetilde{U}(\overline{M})$ is defined to eliminate the radiative
corrections in $U(\overline M)$,
\begin{equation}
\label{eqn:umt}
  \widetilde{U}(\overline{M})\equiv U(\overline{M})/
        \left\{1+\frac{8}{3}\frac{\alpha_\mathrm{s}(\overline{M})}{4\pi}\right \}
\end{equation}
Calculated values of $\widetilde U(\overline M)$ are fitted to the
following parameterisation:
\begin{equation}
\label{eqn:umt_mbar}
  \widetilde{U}(\overline{M})=\omega_0+ \frac{\omega_1}{\overline{M}} 
      + \frac{\omega_2}{\overline{M}^2}
\end{equation}
HQS implies that $\omega_0=1$. However, $\omega_0$ can also be left as
a free parameter to test the applicability of HQS.  Likewise, HQS can
be applied to set $\gamma_P = \gamma_V$ in fits
using equations (\ref{eqn:HQS_fp}) and (\ref{eqn:HQS_fv}). However,
higher order QCD corrections would modify this in a similar way to
$\widetilde{U}(\overline{M})$, that is
\begin{equation}
\label{eqn:HQS_con_rad}
  {\gamma_P \over \gamma_V} = \left(1+\frac{8}{3}
        \frac{\alpha_s(\overline{M})}{4\pi}\right)
\end{equation}
The systematic error arising from extrapolating the decay constants to
the $b$ mass is studied in Section~\ref{sec:decay_const}. The error
from truncating the expansion in inverse powers of the heavy mass is
considered and the issue of the propagation of discretisation effects
under this extrapolation is addressed.

\section{Decay constants}
\label{sec:decay_const}

The main results for the decay constants are listed in
Table~\ref{tab:hl_decay_const} and summarised in the introduction.
Central values are obtained at $\beta=6.2$, setting the scale with
$f_\pi$.  These results are discussed in more detail here, starting
with $f_\mathrm{K}$.

For $f_\mathrm{K}$ the largest source of uncertainty is the choice of
quantity used to set the scale. The value of $f_\mathrm{K}$ increases
by $8\%$ when the scale is set by $r_0$ and decreases by $4\%$ when it
is set by $m_\rho$.  When $m_\phi$ is used to fix the strange quark
mass rather than $m_\mathrm{K}$, the value of $f_\mathrm{K}$ decreases
by $2.5\%$. With only two values of the lattice spacing a continuum
extrapolation is not attempted. The decay constant is smaller at
$\beta=6.0$ by $2.7\%$, which is taken as an estimate of its
discretisation error. These estimates of systematic errors are
combined in quadrature.

\begin{figure}
\begin{center}
\epsfig{file=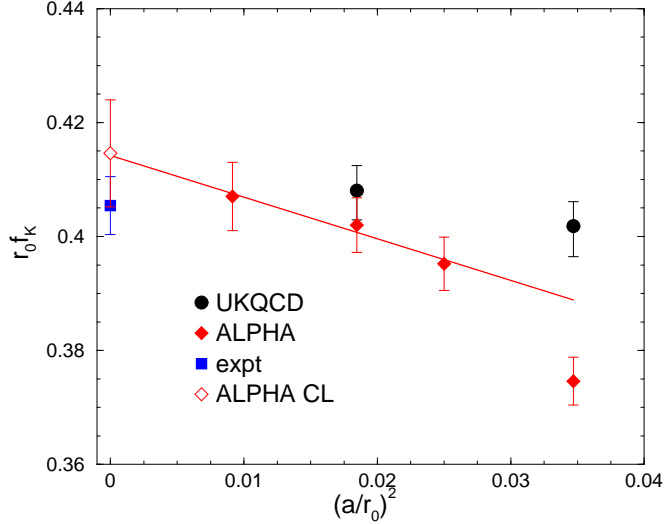,width=0.63\hsize} 
\caption{The dependence of $f_K$ on the lattice spacing.}
\label{fig:fK_comp_ALPHA}
\end{center}
\end{figure}

The ALPHA collaboration \cite{strange_garden} have computed the decay
constant, $f_\mathrm{K}$, with the same action as used here, at
several values of the coupling. They perform a continuum extrapolation
of $r_0f_\mathrm{K}$ against $a^2/r_0^2$. To examine the scaling
behaviour of the decay constant in this work, a comparison with the
ALPHA results is shown in Figure~\ref{fig:fK_comp_ALPHA}. The line
shows the linear extrapolation to the continuum limit (CL) that was
performed by ALPHA, excluding the point at the coarsest lattice
spacing, $\beta=6.0$.  There are two main differences in the
calculations. First, ALPHA use degenerate light quark masses, whereas
this work specifically takes into account the non-degeneracy of the
quarks. However, ALPHA have checked on the same dataset analysised in this
work that using degenerate light quarks has negligible effect. A more
obvious difference is in the values of the improvement coefficients,
$c_A$ and $b_A$. As noted earlier, improvement coefficients may differ
by $\order{a}$ terms, depending on the improvement condition used.  A
particular choice of conditions used to determine the coefficients may
result, for a particular quantity, in smaller discretisation effects
at finite lattice spacing. This seems to be the case for the decay
constants when the improvement coefficients of \Bhatt are used rather
than those determined by ALPHA, especially at $\beta=6.0$. However, as
the continuum limit is approached the values of $f_\mathrm{K}$ from
the two studies converge.

The major source of uncertainty in the heavy-light decay constants
comes from the determination of the lattice spacing.  Additional
systematic errors arise from discretisation effects and from the heavy
quark extrapolations. Since these latter errors are related to the
size of the heavy quark masses, the heavy masses used in this work are
listed in Table~\ref{tab:quark_masses}. All these errors will now be
discussed.
\begin{table}
\begin{center}
\caption{Heavy quark and meson masses used in this work. The table shows
        the {\em bare} quark mass in lattice units, 
        the {\em renormalisation group invariant} quark mass defined by 
        $m^\mathrm{RI}_Q=Z_m(1+b_mam_Q)m_Q$\protect\cite{alpha_ZM} in GeV
        and the heavy-light($\kappa_n$) pseudoscalar meson mass in
        GeV. The scale is set by $f_\pi$.\smallskip}
\label{tab:quark_masses}
\begin{tabular}{cccc}
\hline\hline
\multicolumn{4}{c}{$\beta=6.2$}\\
$\kappa$ & $am_Q$ & ${m}^\mathrm{RI}_Q$ & $M_P$  \\\hline
0.1200  & 0.485    & 1.59 & $2.128(5)$ \\
0.1233  & 0.374   & 1.36 & $1.851(5)$ \\
0.1266  & 0.268   & 1.06 & $1.551(5)$ \\
0.1299  & 0.168   & 0.72 & $1.210(5)$ \\
\hline\hline
\end{tabular}
\hfill
\begin{tabular}{cccc}
\hline\hline
\multicolumn{4}{c}{$\beta=6.0$}\\
$\kappa$ & $am_Q$ & ${m}^\mathrm{RI}_Q$ & $M_P$ \\\hline
  0.1123 & 0.756 & 1.26 & $2.076(8)$\\
  0.1173 & 0.566 & 1.18 & $1.805(6)$ \\
  0.1223 & 0.392 & 0.97 & $1.501(6)$\\
  0.1273 &  0.231 & 0.65 & $1.152(4)$\\
\hline\hline
\end{tabular}
\end{center}
\end{table}

\begin{table}
\begin{center}\caption{The decay constants at both $\beta$ values
        with the scale set by $f_\pi$,$r_0$ and $m_\rho$ and a quadratic (Q) or
        linear (L) fit for the extrapolation in inverse heavy meson
        mass. The quadratic fit uses all four heavy masses, while the
        linear fit uses the heaviest three. The central values are shown
        in bold face.\smallskip}
\label{tab:hl_decay_const}
\begin{tabular}{@{}ll|cccc|cccc@{}}
\hline\hline
 &&\multicolumn{4}{c|}{$\beta=6.2$}  & \multicolumn{4}{c}{$\beta=6.0$} \\
 &&\multicolumn{2}{c}{$f_P$(MeV)}
 &\multicolumn{2}{c|}{$f_V$}
 &\multicolumn{2}{c}{$f_P$(MeV)}
 &\multicolumn{2}{c}{$f_V$}
\\
&&Q&L&Q&L&Q&L&Q&L\\
\hline
&B & ${\bf 195(6)}$ &$178(5)$ &$28(1)$  &$29(1)$
   & $200(8)$ &$182(7)$ &$26(2)$  &$28(2)$\\
&D & ${\bf 206(4)}$ &$207(4)$ &$\ 8.6(3)$ &$\ 8.6(3)$
   & $210(4)$ &$211(5)$ &$\ 8.3(3)$ &$\ 8.3(3)$\\
\raisebox{2ex}[0pt]{$f_\pi$}&$\mathrm{B_s}$
   & ${\bf 220(6)}$ &$201(5)$ &$25.2(8)$  &$26.5(8)$
   & $222(7)$ &$202(7)$ &$25(1)$  &$26(1)$\\
&$\mathrm{D_s}$
   & ${\bf 229(3)}$ &$230(4)$ &$ 8.3(2)$  &$\ 8.3(2)$
   & $230(4)$ &$230(4)$ &$\ 8.3(2)$ &$\ 8.3(2)$\\\hline
&B & $217(5)$ &$200(4)$ &$24.8(8)$  &$25.9(7)$
   & $226(6)$ &$207(5)$ &$18.9(8)$  &$20.4(8)$\\
&D & $222(3)$ &$223(3)$ &$\ 7.8(2)$ &$\ 7.8(2)$
   & $229(3)$ &$230(3)$ &$\ 6.7(2)$ &$\ 6.8(2)$\\
\raisebox{2ex}[0pt]{$r_0$}&$\mathrm{B_s}$
   & $240(4)$ &$222(3)$ &$22.9(5)$  &$24.0(4)$
   & $247(4)$ &$226(3)$ &$21.9(7)$  &$23.1(5)$\\
&$\mathrm{D_s}$
   & $243(2)$ &$244(2)$ &$ 7.7(1)$  &$\ 7.7(1)$
   & $246(2)$ &$247(2)$ &$\ 7.5(1)$ &$\ 7.5(1)$\\\hline
&B & $185(4)$ &$168(3)$ &$29.5(9)$  &$31.1(9)$
   & $197(6)$ &$179(4)$ &$26(1)$ &$28(1)$ \\
&D & $199(3)$ &$199(3)$ &$\ 9.1(2)$ &$\ 9.1(2)$
   & $208(3)$ &$209(3)$ &$\ 8.4(2)$ &$\ 8.4(2)$ \\
\raisebox{2ex}[0pt]{$m_\rho$}&$\mathrm{B_s}$
   & $209(3)$ &$191(2)$ &$26.5(5)$  &$28.0(4)$
   & $220(3)$ &$200(2)$ &$24.9(7)$  &$26.5(5)$ \\
&$\mathrm{D_s}$
   & $221(2)$ &$222(2)$ &$\ 8.7(1)$ &$\ 8.7(1)$
   & $228(2)$ &$228(2)$ &$\ 8.4(1)$ &$\ 8.4(1)$ \\
\hline\hline
\end{tabular}
\end{center}
\end{table}

\begin{figure}
\begin{center}
\epsfig{file=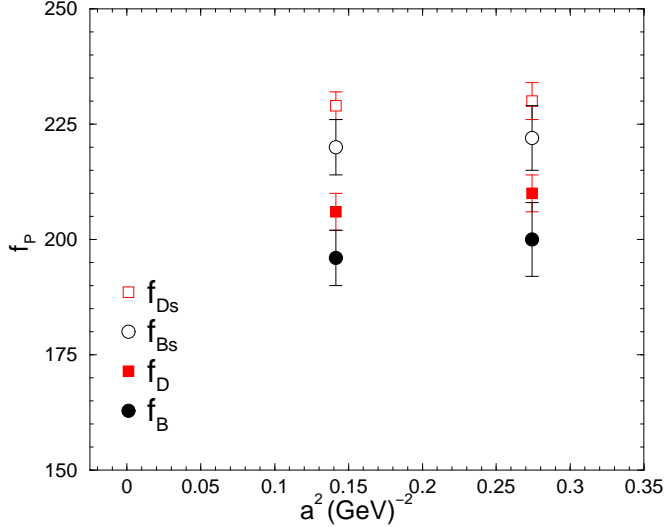,width=0.63\hsize} 
\caption{The dependence of the pseudoscalar decay constants on 
        the lattice spacing. The scale is set by $f_\pi$.}
\label{fig:fB_scale}
\end{center}
\end{figure}

The ambiguity in the decay constants from the lattice spacing is quite
large: there is an overall $16\%$ variation in the value of
$f_\mathrm{B}$ when the scale is set by $m_\rho$, $r_0$ or $f_\pi$
(see Table~~\ref{tab:hl_decay_const}). The difference between the
decay constants determined at $\beta=6.2$ and $\beta=6.0$ is smallest
when the scale is set by $f_\pi$, as shown in
Figure~\ref{fig:fB_scale}. This is not surprising since some
systematic uncertainties may cancel in ratios of decay
constants. Therefore the central values are quoted using $f_\pi$ to
set the scale.

Discretisation errors are considered next. Although this calculation
is $\order{a}$ improved, $\order{a^2}$ effects may still be important
at any fixed lattice spacing. In particular, $\order{a^2 m^2}$ effects
could well be significant for heavy quarks. With results at only two
values of the lattice spacing, a full continuum extrapolation cannot
be attempted. The heavy-light decay constant at the finer lattice
spacing is taken as the central value, with the result at the coarser
spacing used as one indicator of discretisation errors. A fuller
investigation of discretisation effects for the heavy quarks follows.

\begin{figure}
\epsfig{file=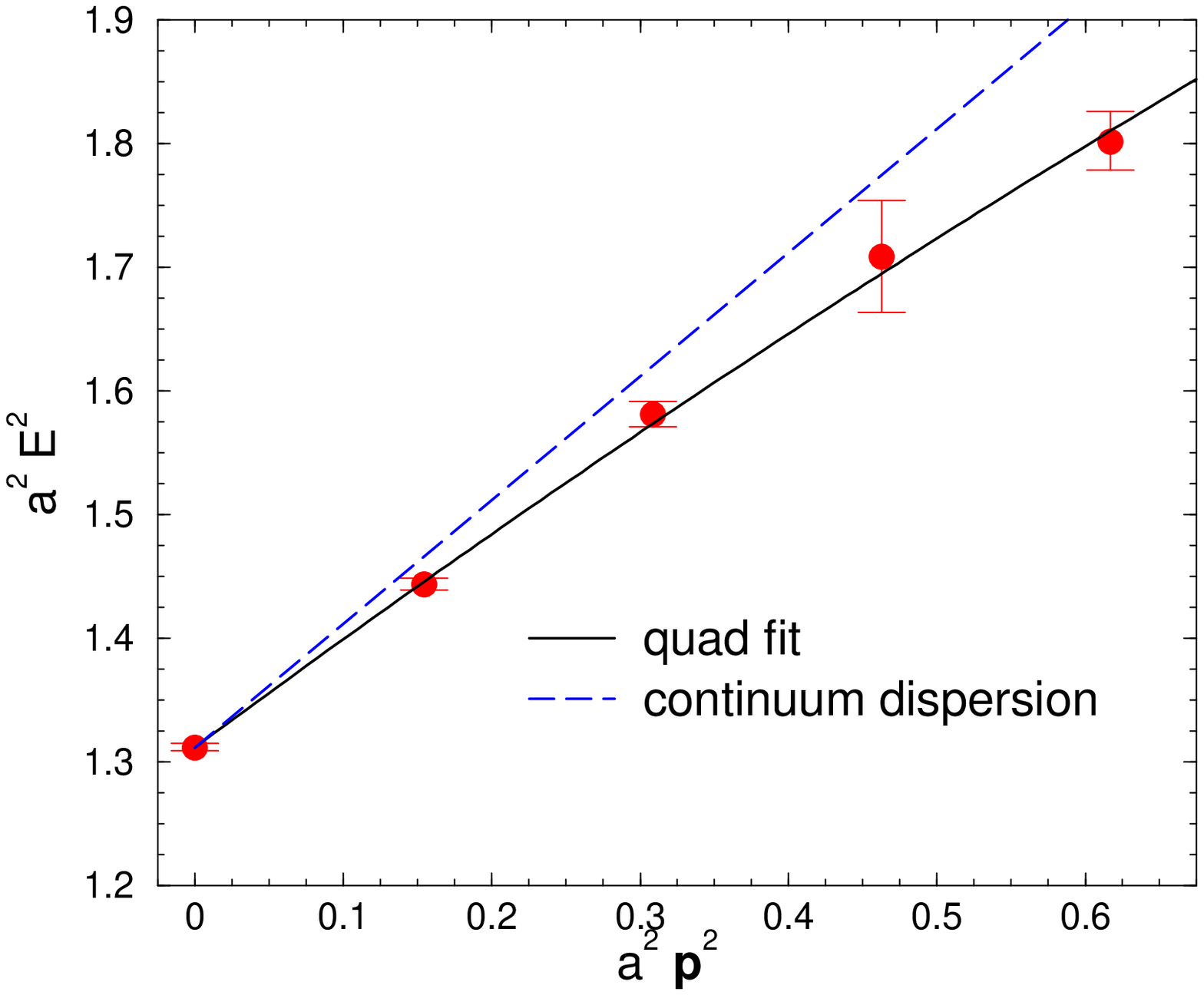,width=0.46\hsize}
\hfill
\epsfig{file=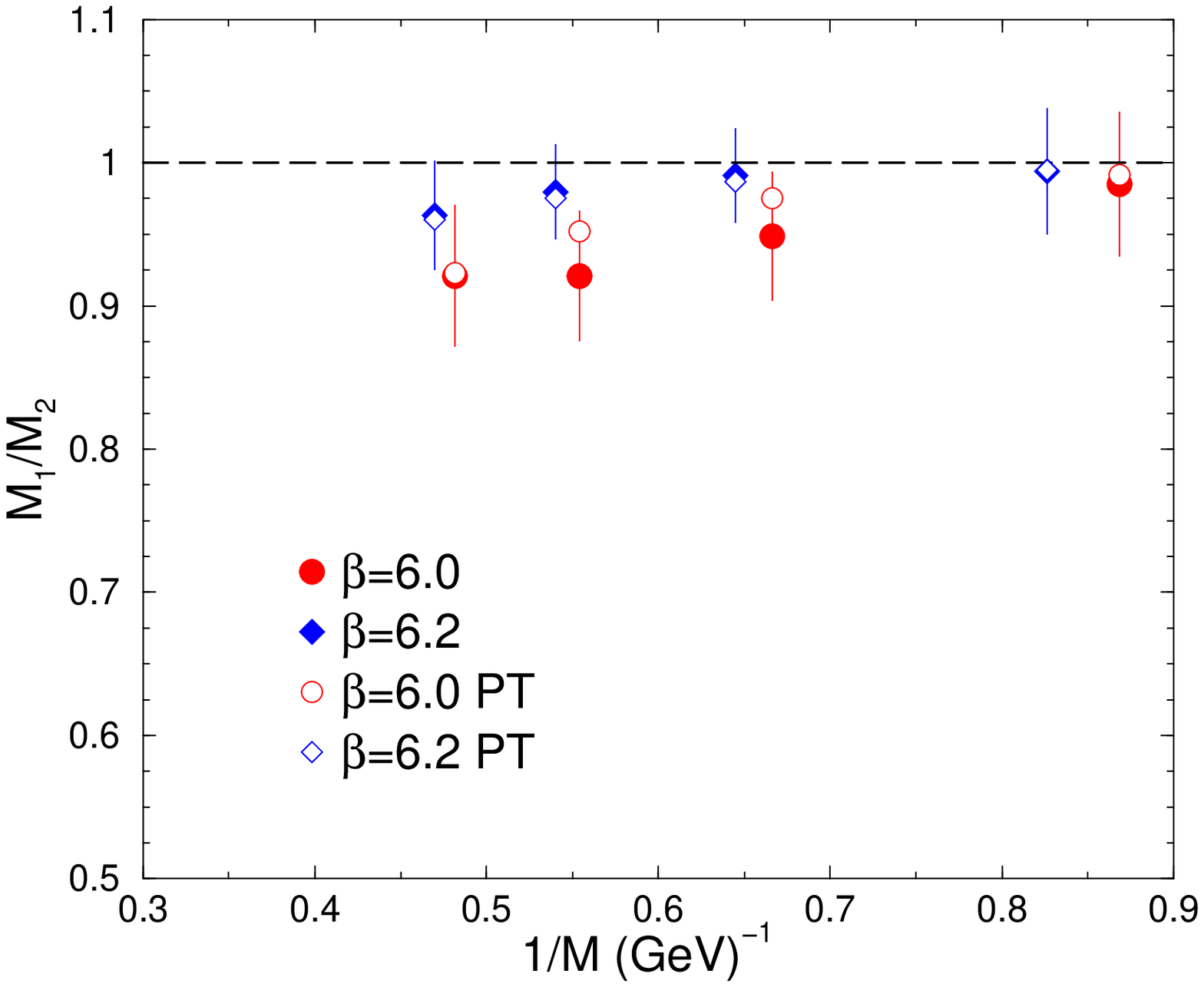,width=0.46\hsize}  
\caption{Left: the pseudoscalar dispersion relation for $\kappa_H=0.1123$,
        $\kappa_L=0.13344$ in lattice units at $\beta=6.0$. Right: the
        ratio $M_1/M_2$ as a function of $M_1$.}
\label{fig:disp_rel-m1m2}
\end{figure}

The general characteristic size of discretisation effects in this
simulation is investigated by examining the free particle dispersion
relation, which is altered in discrete spacetime. The lattice
dispersion relation can be written
\begin{equation}
  E^2 = M_1^2 + \frac{M_1}{M_2}\vec{p}^{\ 2} + {\mathcal{O}}(p^4)
\end{equation}
where $M_1$ is the energy at zero momentum and $M_2$ is the kinetic
mass, defined by $M_2^{-1}=\left.\partial^2 E/\partial p_i^2 \right
|_{\vec{p}=0}$.  The dispersion relation is investigated by fitting
hadronic correlators computed at five different momentum values
($\vec{p}^{\;2}=0,1,2,3,4$ in lattice units)
using linear and quadratic fits in $\vec{p}^{\;2}$. The fit for the
heaviest quark combination at $\beta=6.0$ is shown on the left of
Figure~\ref{fig:disp_rel-m1m2}. For lighter mass combinations at
$\beta=6.0$ the signal for highest momentum is very poor, and so in
general the fits exclude the $|\vec{p\;}|^2=4$ channel. The right hand side
of Figure~\ref{fig:disp_rel-m1m2} shows the $M_1/M_2$ estimates for
each heavy quark.
Also shown on the right in Figure \ref{fig:disp_rel-m1m2} is the
ratio of $M_1/M_2$ when $M_2$ has been determined by the shift in quark
mass from $m_1$ to $m_2$~\cite{bernard_fB_old}.
\begin{equation}
  M_2^\mathrm{PT}=M_1 + (m_2-m_1)
\end{equation}
and the quark masses $m_1$ and $m_2$ have been calculated from the 
tree-level relation described in ~\cite{KLM_norm,L_norm,K_norm}.
\begin{equation}
  am_1=\ln(1+am_Q)
\end{equation}
\begin{equation}
\label{eqn:EKM}
  \frac{1}{am_2}=\frac{2}{am_Q(2+am_Q)}+\frac{1}{1+am_Q}
\end{equation}
Here $M$ denotes hadron mass and $m$ denotes quark mass. There is good
agreement between the tree-level perturbative description of $M_2$ and
the NP determination. At the quark masses used in this work the ratio
$M_1/M_2$ remains close to unity, so that deviations from the
continuum dispersion relation are small. Finally, it is worth noting
that $M_2$ is determined at non-zero momentum, which is statistically
noiser, and so has statistical errors an order of magnitude larger
than for $M_1$.

The axial current in this work is normalised according to
equations~(\ref{eqn:imp_current},\ref{eqn:renorm_current}). As already
mentioned the improvement term is proportional to $\sinh(aM_P)$ and
the mass-dependent normalisation is proportional to $am_Q$.
Bernard~\cite{Bernard_2000} proposed the following alternative
normalisation, labelled CB,
\begin{equation}
\label{eqn:CB_norm}
  (A_0^R)_\mathrm{CB}=Z_A  \left [1 + b_A a\bar{m} + 2 a c_A 
     \frac{\partial_0 \langle 0 | P | P \rangle}
     {\bar{m}\langle 0 | A_0 | P \rangle } \right ]^{1/2} A_0
\end{equation}
where $\bar{m}$ is the average of the bare quark masses in the
heavy-light state.  The CB normalisation differs from the NP
improvement scheme normalisation only at $\order{a^2}$. Taking into
account an additional normalisation of $\sqrt{4\kappa_Q\kappa_q}$, the
CB norm has a finite static limit since $\kappa_Q a m_Q \to 1/2$ in
as $\kappa_Q \to 0$.  The CB norm looks very like the
normalisation of Kronfeld, Lepage and Mackenzie (KLM
norm)~\cite{KLM_norm,lepage_mack,L_norm,K_norm},
\begin{equation}
  \psi \to \psi^{\prime}=\psi \sqrt{1+\mu am}
\end{equation}
where $\mu$ would be the mean link in the tadpole improved tree-level
Fermilab formalism. In this case $\mu$ is a mass-dependent factor that
contains the information about improving the current.

\begin{figure}
\begin{center}
\epsfig{file=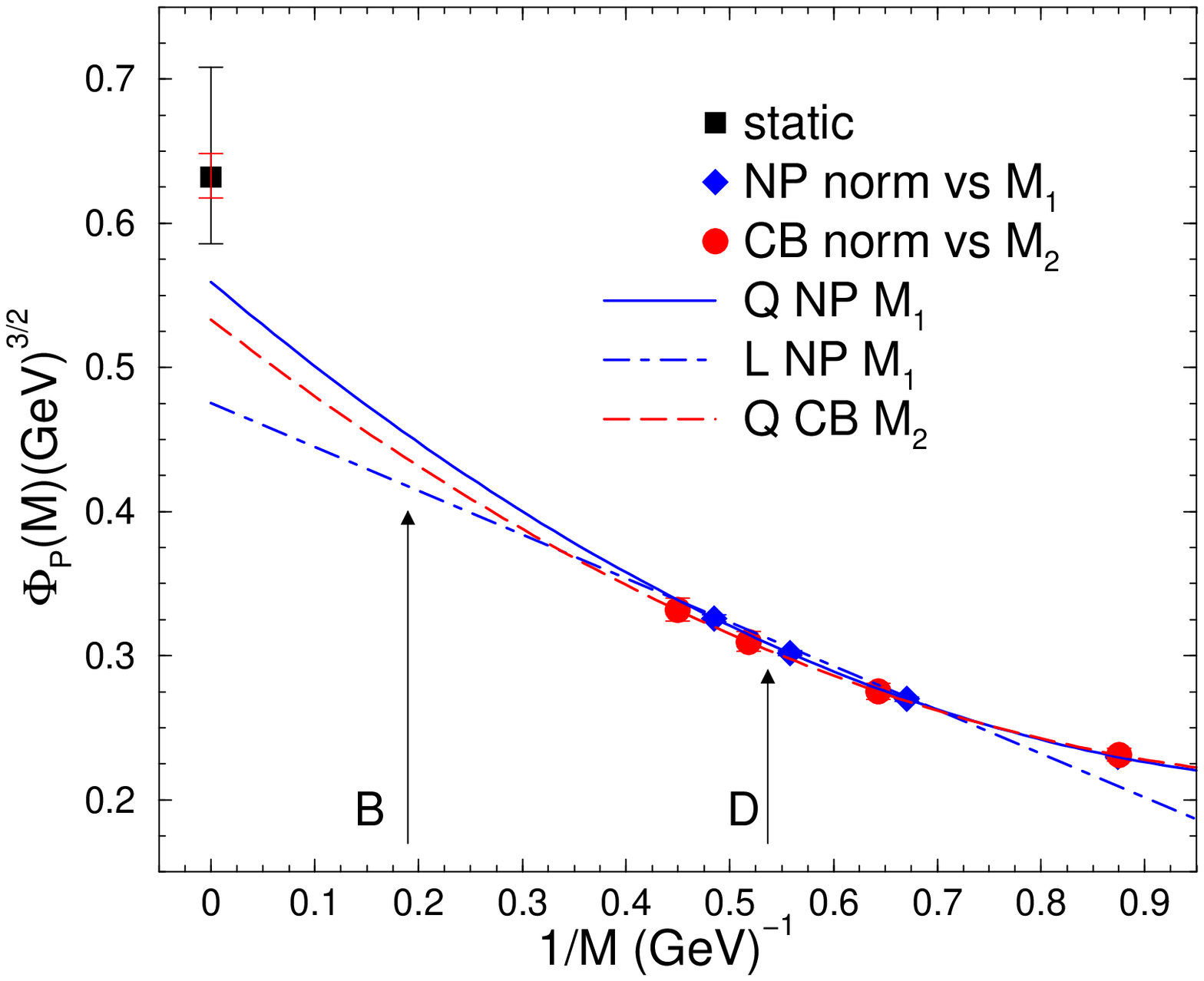,width=0.46\hsize}
\hfill 
\epsfig{file=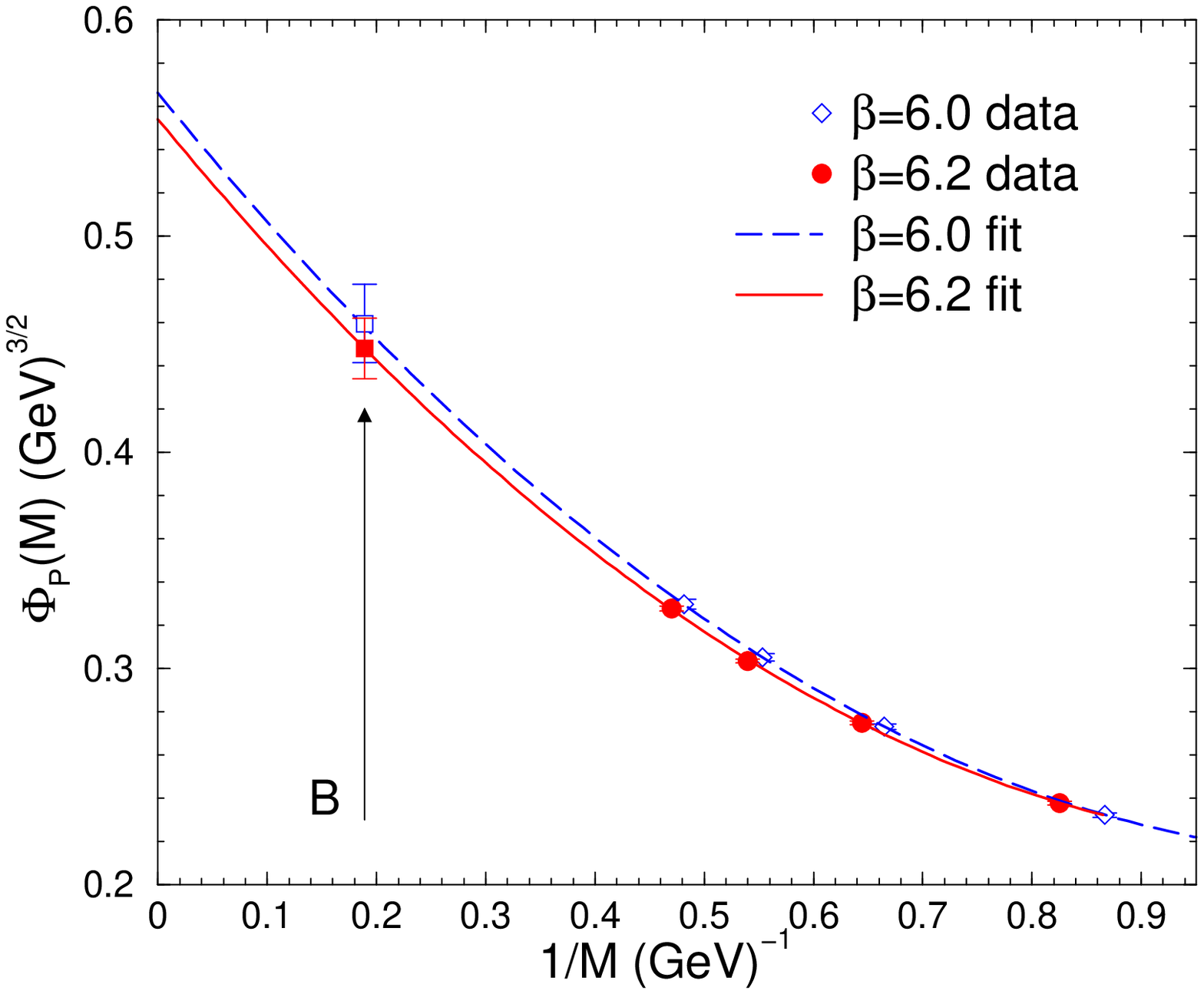,width=0.46\hsize} 
\caption{The left plot shows the extrapolations of the pseudoscalar decay 
        constant in heavy meson mass at $\beta=6.0$ with different 
        normalisations and masses. The right plot shows a comparison between
        the extrapolations at both values of the coupling.}
\label{fig:phi_60}
\end{center}
\end{figure}

Figure~\ref{fig:phi_60} shows the mass dependence of the function
$\Phi$, defined in equation (\ref{eqn:HQS_fp}), for a variety of
methodologies. The NP norm is shown with both a quadratic fit (Q) to
all four masses and a linear (L) fit to the heaviest three. The CB
norm data has the extrapolation using the $M_2$ masses rather than
$M_1$.  It is clear from the figure that the difference between the
``NP Q vs $M_1$'' and ``CB Q vs $M_2$'' fits is smaller than the
difference between the quadratic and linear fits to NP norm.  It
should be noted that the effect of changing the normalisation to CB
differs from that described in \cite{Bernard_2000} which used the
ALPHA determination of $c_A$ and the preliminary NP determination by
\Bhatt~\cite{bhatta_plb} of $b_A$.

In the $\order{a}$ improved theory used here, $\order{a^2m_Q^2}$
lattice artefacts could still be large: the worst case simulated has
$a^2m_Q^2\sim 0.5$ (see Table~\ref{tab:quark_masses}) and one
may worry that mass-dependent discretisation effects, which may not be
too large at the charm scale, are enhanced by extrapolation to the
bottom scale. Although a Taylor expansion of $(am_Q)^2$ in $1/m_Q$ around the
charm quark mass needs large terms to reproduce $(am_b)^2$,
the issue is not directly the fate of $\order{a^2m_Q^2}$ under
extrapolation, but rather the effect of such terms at the charm scale
on the extrapolation.

The left hand plot in Figure~\ref{fig:phi_60} includes the static
point. The smaller set of error bars shows the error from
renormalisation, the larger, the statistical uncertainty. The static
point is higher than the extrapolations; however, the missing static
improvement term would lower the static point, since
$c_A^\mathrm{stat}$ is negative. For this reason an interpolation to
the $b$ quark mass using the static and $c$ quark data is not
implemented; rather the static point is used as a check of the
extrapolation.  The result supports the view that, the
increase in discretisation error after the heavy quark extrapolation
is moderate.

\begin{figure}
\begin{center}
\epsfig{file=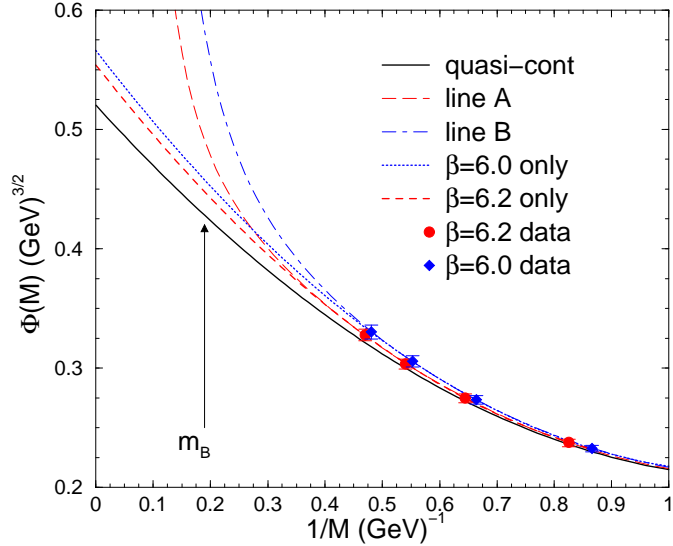,width=0.63\hsize}
\caption{Effects of lattice artefacts on the heavy quark extrapolation.
         The solid line shows the quasi-continuum extrapolation
         described in the text. The dashed and dotted lines show
         $\Phi$, equation~(\ref{eqn:HQS_fp}), when extrapolating at
         each value of $\beta$ separately. Line A(B) shows the
         function $\Phi(M,a)$, equation~(\ref{eqn:quasi_cont_phi}), at
         $\beta=6.2(6.0)$.}
\label{fig:quasi_cont_phi}
\end{center}
\end{figure}

The right-hand side of Figure~\ref{fig:phi_60} shows the extrapolation
of $\Phi_P$ for both values of $\beta$ in physical units. Although the
difference between the curves does indeed grow during the
extrapolation, it is smaller than the statistical errors at the bottom
scale. The qualitative agreement between the two curves suggests that
discretisation errors are not large.  To study this further,
equation~(\ref{eqn:HQS_fp}) is modified as follows,
\begin{equation}
\label{eqn:quasi_cont_phi}
\Phi(M,a) = \gamma\left(1 + \frac{\delta}{M} + \frac{\eta}{M^2} 
         + \varepsilon (a M )^2 + \zeta (a M)^3 \right)
\end{equation}
and data from both lattice spacings fitted simultaneously. This uses
the two values of $a$ to increase the number of $aM$ values available,
allowing the important class of heavy-mass dependent discretisation
errors to be studied. The sum of the first three terms in
equation~(\ref{eqn:quasi_cont_phi}) has these errors subtracted out
and is hereafter referred to as the `quasi-continuum'
result. Figure~\ref{fig:quasi_cont_phi} shows the quasi-continuum
curve as a function of $1/M$, together with $\Phi(M,a)$ at each
$\beta$ value separately. The curves A and B which blow up as $1/M \to
0$ are the fits to equation~(\ref{eqn:quasi_cont_phi}) containing
lattice artefacts. However, the quasi-continuum result, where these
artefacts are subtracted, does not differ greatly from the
extrapolations at the individual lattice spacings using
equation~(\ref{eqn:HQS_fp}).  The quasi-continuum $\Phi$ gives
$f_B=186(10)$ MeV, which differs by $5\%$ at $\beta=6.2$ (the number
appearing in Table~\ref{tab:sys_error}) and $7\%$ at $\beta=6.0$ from
the extrapolations using~(\ref{eqn:HQS_fp}).  Excluding the $\zeta$
term, or using $am_Q$ instead of $aM$ makes little difference. Using
$(aM)^2 \Lambda/M$ instead of $(aM)^2$ also makes no significant
difference. Varying all five fit parameters in
equation~(\ref{eqn:quasi_cont_phi}) over the region where the
chi-squared per degree of freedom increases by up to $1$ from its
minimum value gives a variation of $\pm 16$ MeV for the
quasi-continuum $f_B$, so the fit is stable.

The systematic error associated with truncating the HQET series is now
discussed. First, linear and quadratic fits to the data are
compared. The linear fit uses the heaviest three quark masses and
drops the $\eta$ term from equations~(\ref{eqn:HQS_fp}) and
(\ref{eqn:HQS_fv}). This is compared to a quadratic fit using all four
masses. The resulting variations, which are shown in
Table~\ref{tab:hl_decay_const}, amount to a relative error of $9\%$
for $f_\mathrm{B}$, indicating that a quadratic term is necessary. A
cubic fit is performed to check that a quadratic term is sufficient.
There are not enough data points for a cubic fit at each lattice
spacing, so the fit combines data from both lattices as before, adding
a cubic term $\xi/M^3$ to $\Phi(M,a)$. This results in a value of
$f_\mathrm{B}$ which is $5\%$ greater than the quasi-continuum
determination.  This is also included as a systematic error in
Table~\ref{tab:sys_error}.

\begin{figure}
\begin{center}
\epsfig{file=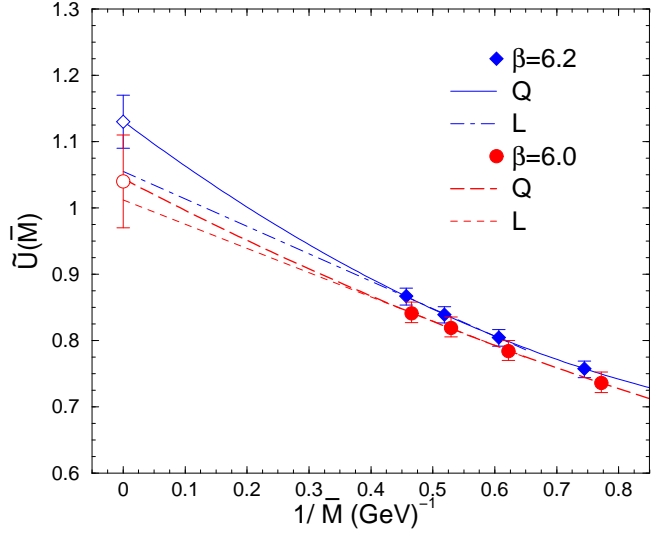,width=0.63\hsize} 
\caption{The quantity $\widetilde{U}(\overline{M})$ as a function of the
        inverse spin averaged mass $\overline{M}$.}
\label{fig:UM}
\end{center}
\end{figure}

The HQS relation between the pseudoscalar and vector decay constants
can also be investigated. The quantity $\widetilde{U}(\overline{M})$
defined in equation~(\ref{eqn:umt}) should be equal to unity in the
static limit.  The extrapolation of $\widetilde{U}(\overline{M})$ is
shown in Figure~\ref{fig:UM} and displayed in
Table~\ref{tab:UM}. ``Q'' denotes a quadratic fit to all heavy quarks
whereas ``L'' denotes a linear fit to the heaviest three.  The
extrapolation for $\beta=6.0$ displays the expected static limit,
albeit with large errors. At $\beta=6.2$,
$\widetilde{U}(\overline{M})$ from the quadratic fit deviates from
unity in the static limit. On the finer lattice, discretisation errors
are smaller and so one might expect better agreement in the static
limit.  However, $\widetilde{U}(\overline{M})$ depends on the ratio of
the axial to vector currents and is particularly sensitive to the
mixing coefficients, $c_A$ and $c_V$, which are the most poorly known
of the improvement coefficients.  Different determinations of
$c_V$ vary by an order of magnitude, even at $\beta=6.2$,
where all the other coefficients are in much better agreement. Indeed,
varying $c_V$ by the quoted errors of $25\%$ in \Bhatt varies the
value of $\widetilde{U}(\overline{M}_\infty)$ by around $6\%$. Taking
into account the uncertainty in the value of $c_V$, the value of
$\widetilde{U}(\overline{M}_\infty)$ is consistent with unity.
\begin{table}
\begin{center}
\caption{$\widetilde{U}(\overline{M})$ as a function of heavy meson mass.
        \smallskip}
\label{tab:UM}
\begin{tabular}{ccccc}\hline\hline
$\beta$&& $ \overline{M}_\mathrm{D}$& $ \overline{M}_\mathrm{B}$&
  $ \overline{M}_\infty$\\\hline
      &Q & $0.83(1) $& $1.01(2)$& $1.13(4)$\\
\middle{$6.2$} &L & $0.83(1)$ & $0.98(2)$& $1.05(5)$ \\
 &Q & $0.82(2) $& $0.96(4)$& $1.04(7)$\\
\middle{$6.0$} &L & $0.82(2) $& $0.94(3)$& $1.01(4)$\\\hline\hline
\end{tabular}
\end{center}
\end{table}

The $SU(3)$ flavour breaking ratios are determined by two methods:
from the ratio of extrapolated values of the decay constants, labelled
(i) in Table~\ref{tab:flavour}, or by extrapolating the ratios
constructed at each heavy mass, labelled (ii) in the Table. In the
ratio of $f_{P_\mathrm{s}}/f_{P_\mathrm{n}}$ most of the heavy quark mass
dependence seems to cancel. The fact that the two methods agree,
within statistical errors, suggests that the systematic error in these
ratios is small.

\begin{table}
\begin{center}
\caption{Flavour-breaking ratios.\smallskip}
\label{tab:flavour}
\begin{tabular}{cccccc}\hline\hline
$\beta$ &&  $\frac{f_{\mathrm{B_s}}}{f_{\mathrm{B}}}$
        &  $\frac{f_{\mathrm{D_s}}}{f_{\mathrm{D}}}$
        &  $\frac{f_{\mathrm{B^{\ast}}}}{f_{\mathrm{B_s^{\ast}}}}$
        &  $\frac{f_{\mathrm{D^{\ast}}}}{f_{\mathrm{D_s^{\ast}}}}$
\\\hline
 &(i)&$1.13(1)$&$1.11(1)$&$1.10(2)$&$1.04(1)$\\
\middle{$6.2$}&(ii)&$1.11(2)$&$1.11(1)$&$1.10(2)$&$1.05(1)$\\
 &(i)&$1.11(2)$&$1.09(1)$&$1.05(3)$&$1.00(2)$\\
\middle{$6.0$}&(ii)&$1.10(2)$&$1.09(1)$&$1.05(3)$&$1.01(2)$\\\hline\hline
\end{tabular}
\end{center}
\end{table}
The systematic variations of the decay constants are shown in 
Table~\ref{tab:sys_error}, in order of importance. The total systematic 
uncertainty is obtained by combining these in quadrature.

\begin{table}
\begin{center}
\caption{Percentage systematic uncertainties. Systematic differences
are obtained by varying the procedure used to calculate the decay
constants. The central values are fixed by using $f_\pi$ to set the
scale at $\beta=6.2$, performing a quadratic heavy quark
extrapolation, taking the central values of the mixing coefficients
$c_{A,V}$, fitting correlation functions to a single exponential, and
using $m_{\mathrm K}^2$ to set the strange quark mass.}
\label{tab:sys_error}
\smallskip
\begin{tabular}{crrrrrrr}\hline\hline
Pseudoscalar &$f_{\mathrm{B}}$   
&$f_{\mathrm{D}}$ 
&$f_{\mathrm{B_s}}$ 
&$f_{\mathrm{D_s}}$ 
&$\frac{f_{\mathrm{B_s}}}{f_{\mathrm{B}}}$   
&$\frac{f_{\mathrm{D_s}}}{f_{\mathrm{D}}}$    \\ \hline
scale set by $r_0$    & $11$ & $8$  & $9$  & $6$  & $-2$ & $-2$\\
scale set by $m_\rho$ & $-5$ & $-3$ & $-5$ & $-3$ & $1$  & $1$\\
linear vs quadratic   & $-9$ & $-$  & $-9$ & $-$  & $-$  & $-$ \\
quasi-continuum       & $-5$ & $-2$ & $-5$ & $-2$ & $-$  & $-$ \\
quasi-continuum plus $1/M^3$
                      & $5 $&$-$ &$ 5 $& $-$ & $-$  & $-$ \\
$\beta=6.0$           & $3 $&$ 2 $&$ 1 $&$ 1 $&$ -1 $&$ -2$ \\
multi-exp             & $-2$&$-3$&$-5$&$-3$&$-3$&$1$\\
strange quark mass from $m_\phi$
                      & $-$& $-$&$-2$&$-2$&$ -2$&$-2$\\
coeff. $c_{\mathrm{A}}$ 
                      & $1 $&$ 1$&$ 1 $&$ 1 $& $1$ & $-$ \\\hline
Vector&$f_{\mathrm{B^\ast}}$ &$f_{\mathrm{D^\ast}}$ &$f_{\mathrm{B_s^\ast}}$   
&$f_{\mathrm{D_s^\ast}}$ 
&$\frac{f_\mathrm{B^*}}{f_\mathrm{B^*_s}}$ 
&$\frac{f_\mathrm{D^*}}{f_\mathrm{D^*_s}}$ \\\hline
scale set by $r_0$ & $-11$&$ -10$&$ -9$&$ -8$&$ 2$&$    2$\\ 
scale set by $m_\rho$ &$ 6$&$ 5$&$ 5$&$ 5$&$ 1$&$    1$\\ 
$\beta=6.0$ &$ -7 $&$ -4 $&$ -2 $&$ -1 $&$ -5 $&$ -4$ \\
linear vs quadratic & $5$& $-$ &$ 5$& $-$ & $-$& $-$ \\ 
quasi-continuum       & $6$ & $3$ & $6$ & $3$ & $-$  & $-$ \\
quasi-continuum plus $1/M^3$
                      & $-5$ & $-$ & $-5$ & $-$ & $-$  & $-$ \\
strange quark mass & $-$& $-$& $ 2$&$  2$& $-3$ &$-2$\\
coeff $c_V$ & $3$& $1$&  $3$&  $1$& $1$ &$1$\\
\hline\hline
\end{tabular}
\end{center}
\end{table}

\subsection{Comparison with other determinations}

A comparison with some other recent quenched results for $f_\mathrm{B}$
is shown in Table~\ref{tab:fB_comp}. For a comprehensive review, the
reader is referred to the article by Bernard~\cite{Bernard_2000}. The
results which use the Fermilab formalism~\cite{KLM_norm} are continuum
limits. The APE result uses the same action at $\beta=6.2$ and the
UKQCD Tad result uses the same gauge configurations as this work but
with a tadpole improved value of $c_{\mathrm{SW}}$.

\begin{table}
\begin{center}
\caption{Comparison with some other recent determinations of $f_\mathrm{B}$. 
  ``FNAL'' denotes use of the Fermilab formalism~\cite{KLM_norm} and 
  ``charm'' indicates that  the heavy quarks
  have masses around charm and are then extrapolated to the bottom scale.
  NB: the World Average is not the average of the numbers
  displayed below.\smallskip}
\label{tab:fB_comp}
\begin{tabular}{ccl}
\hline\hline
 & Heavy Quark & $f_{\mathrm B}$ (MeV)  \\\hline
El-Khadra {\em et al.}~\cite{KLM_decay} & FNAL & $164(8)\pme{11}{\ 8}$  \\
MILC~\cite{MILC_fb,MILC_2000_fB} & FNAL  & $173(6)\pme{16}{16}$  \\
JLQCD~\cite{JLQCD_fB}& FNAL  & $173(4)(9)(9)$  \\
UKQCD Tad~\cite{dlin_Bmix} & charm & $177(17)\pme{22}{26}$\\
APE~\cite{APE_fB_new} & charm & $179(18)\pme{34}{\ 9}$  \\
CP-PACS~\cite{CP-PACS_fB} & FNAL & $188(3)(9)$\\
This Work & charm & $195(6)\pme{24}{23}$  \\\hline
World Average~\cite{Bernard_2000} & &$175(20)$\\\hline\hline 
\end{tabular}
\end{center}
\end{table}

The value of $f_\mathrm{B}$ obtained in this work is the highest,
although it is compatible with the world average. One reason for this
might be the heavy quark method employed. However, other results using
the same heavy quark extrapolation have a lower value, while the most
recent result using the Fermilab method from the CP-PACS
collaboration~\cite{CP-PACS_fB} has a value in agreement within
statistical errors. Different heavy quark methods have different
associated systematic uncertainties. A detailed analysis of those
present in this work has been discussed above, including an estimate
of the effects of the heavy quark extrapolation. These uncertainties
are also explored in~\cite{Bernard_2000}, which additionally addresses
some of the systematic issues affecting the Fermilab formalism. The
value obtained in this work is consistent within these systematic
uncertainties with other quenched results.

There have been several recent calculations of $f_\mathrm{B}$ with two
flavours ($N_F=2$) of dynamical fermions using NRQCD or the
Fermilab formalism. Again, they are reviewed in~\cite{Bernard_2000}. The
effect of unquenching is found to increase the value of $f_\mathrm{B}$ by
$10-15\%$.

The only heavy-light decay constant to have been measured experimentally so 
far is $f_\mathrm{D_s}$~\cite{fD_franz},
\begin{equation}
  f_\mathrm{D_s}=260\pm19\pm32 \ \mathrm{MeV}
\end{equation}
With such large uncertainties, the experimental value is consistent with
both the unquenched and quenched values of $f_\mathrm{D_s}$.

\section{Spectroscopic quantities}
\label{sec:spectro}

To determine the value of the hopping parameter, $\kappa_c$,
corresponding to the charm quark mass requires an interpolation in
heavy quark mass.  The improved quark mass definition in
equation~(\ref{eqn:mimp}) also applies to heavy quarks, so if the
meson mass depends linearly on the improved heavy quark mass, the
corresponding \emph{bare} quark mass dependence is,
\begin{equation}
\label{eqn:bare_amQ}
  a M_\mathrm{H}(m_Q)= \rho + \lambda am_Q + \epsilon a^2m_Q^2
\end{equation}
where $\epsilon/\lambda=b_m$. In the NP improved formulation
all lattice artefacts of ${\mathcal{O}}(a)$ have been
removed. However, ${\mathcal{O}}(a^2m_Q^2)$ effects for the heaviest
quarks could be significant. Any contributions at this order would
affect the ratio $\epsilon/\lambda$ such that it was no longer equal
to $b_m$.  A fit to equation (\ref{eqn:bare_amQ}) was tried with
$\epsilon/\lambda$ fixed to $b_m$ from boosted perturbation
theory, and with $\epsilon/\lambda$ allowed to vary
freely\footnote{This procedure is entirely equivalent to that of
Becirevic {\em et al.}~\cite{APE_fB_new}.}.  The heavy quark
dependence can be used to fix $\kappa_\mathrm{c}$ by choosing a
particular state (or splitting) to have its physical value.  Choosing
the pseudoscalar mass to fix $\kappa_\mathrm{c}$, the spectrum of
heavy-light mesons can then be predicted.

\begin{figure}
\begin{center}
\epsfig{file=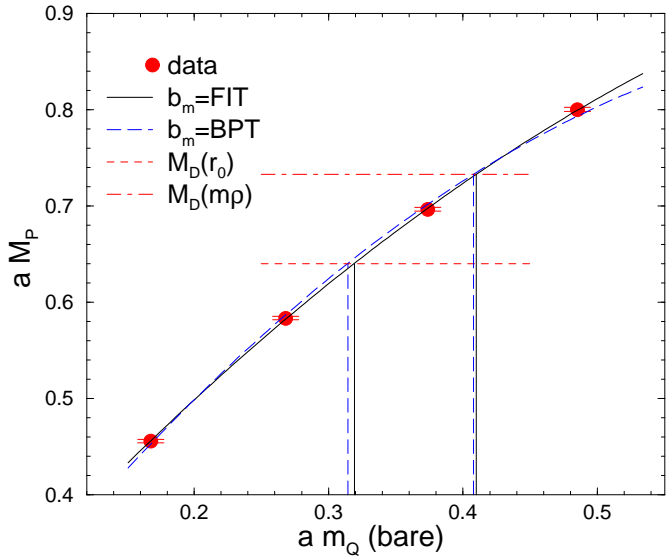,width=0.63\hsize} 
\caption{The pseudoscalar meson vs. bare quark mass for $\beta=6.2$}
\label{fig:charm_62}
\end{center}
\end{figure}

For $\beta=6.2$, the results of using equation~(\ref{eqn:bare_amQ})
are shown in Figure \ref{fig:charm_62}. Whilst the value of
$\epsilon/\lambda=-0.505(4)$ (labelled FIT in the figure) differs
somewhat from the value of $b_m=-0.652$ from BPT, it makes
little difference to the value of the $\kappa_\mathrm{c}$, of order
$0.1\%$. The choice of quantity to set the lattice spacing clearly has
a rather large effect.  Values for $\kappa_\mathrm{c}$ are shown in
Table~\ref{tab:kcharm}, using the free fit to set
the value of the hopping parameter.
\begin{table}
\begin{center}
\caption{The hopping parameter corresponding to the charm quark mass.
        The label FIT refers to the ratio $\epsilon/\lambda$ in
        equation (\ref{eqn:bare_amQ}) having a fitted value, whilst BPT refers
        to the ratio having the BPT value.\smallskip}
\label{tab:kcharm}
\begin{tabular}{ccccc}\hline\hline
 & \multicolumn{2}{c}{$\beta=6.2$} & 
        \multicolumn{2}{c}{$\beta=6.0$} \\
$a^{-1}$ & FIT & BPT & FIT & BPT \\\hline
$a^{-1}(r_0)$ & $0.12498^{+6}_{-5}$ & $0.12513^{+6}_{-5}$ &
        $0.11952^{+8}_{-6}$ & $0.12056^{+8}_{-6}$ \\    
$a^{-1}(m_\rho)$ & $0.12221^{+7}_{-5}$ & $0.12227^{+7}_{-5}$ &
        $0.1160^{+1}_{-1}$ & $0.1165^{+2}_{-2}$ \\
$a^{-1}(f_\pi)$ & $0.1232^{+6}_{-5}$  & $0.1233^{+6}_{-6}$ &
        $0.1164^{+10}_{-11}$ & $0.1171^{+3}_{-10}$ \\\hline\hline
\end{tabular}
\end{center}
\end{table}

For $\beta=6.0$, the fitted value of $\epsilon/\lambda=-0.384(3)$
differs significantly from the value of $b_m=-0.662$ from
BPT. Consequently the fit using the BPT $b_m$ has a very
large $\chi^2$, as the data cannot accomodate a model with such large
curvature. However, the difference in $\kappa_\mathrm{c}$ is still
small (of order $1\%$). For $\beta=6.2$ the heaviest quark has a value
of $am_Q=0.485$ whereas for $\beta=6.0$ the heaviest quark has
$am_Q=0.775$.  Discretisation errors of $\order{a^2m^2}$ could be
responsible for modifying the value of the ratio
$\epsilon/\lambda$. The value of $\kappa_\mathrm{c}$ is rather
insensitive to $\epsilon/\lambda$ but because the value of
$am_\mathrm{Q}$ is so large, the improved quark mass changes
dramatically. Using the free fit as the preferred method, the value of
$\kappa_\mathrm{c}$ is shown in Table~\ref{tab:kcharm}.

\begin{figure}
\begin{center}
\epsfig{file=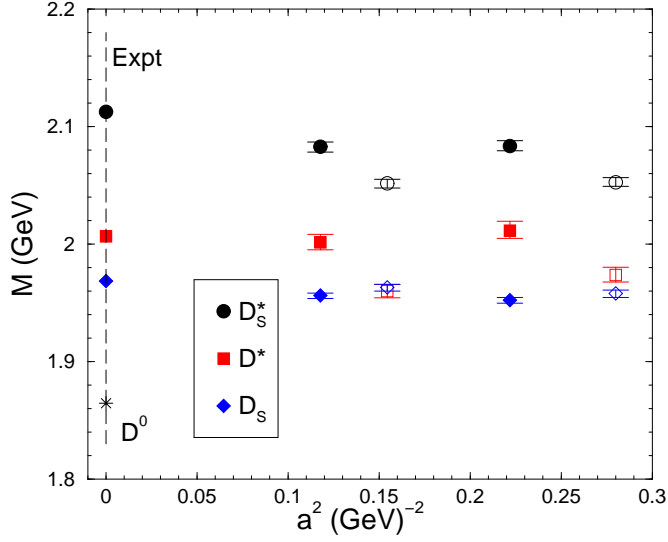,width=0.63\hsize} 
\caption{The scaling of the D meson spectrum. The filled symbols have
        the lattice spacing set by $r_0$, and the open symbols by
        $m_{\rho}$.}
\label{fig:D_spec}
\end{center}
\end{figure}

The spectrum of heavy-light charm states is shown in
Figure~\ref{fig:D_spec} and tabulated in Table~\ref{tab:D_spec}. With
only two values of $\beta$ a continuum extrapolation is not attempted,
but the difference between the two couplings can be used to estimate
systematic errors. It is clear from the figure that a large systematic
error arises from the definition of the lattice spacing. Depending on
which quantity is chosen, different mass splittings appear to be
closer to their continuum values.  The central values for the spectrum
are produced by the following procedures: single exponential fits,
using $r_0$ to set the scale, with $\epsilon/\lambda$ a free parameter
in equation~(\ref{eqn:bare_amQ}). The values of the masses at
$\beta=6.2$ are taken to be the central values, with the values at
$\beta=6.0$ used to estimate systematic error. The remaining estimates
of systematic error come from using multiple exponential fits, using
$m_{\rho}$ to set the scale and, in the case of states with a strange
quark, using $m_{\mathrm K^*}$ instead of $m_{\mathrm K}^2$ to set the
strange quark mass. All these systematic errors are combined in
quadrature. The results, with experimental comparisons, are:
\begin{equation}
\begin{array}{rcl}
\multicolumn{3}{c}{\mbox{Lattice, this work}}\\
  M_{\mathrm D_{\mathrm s}} &=&
    1.956^{+2}_{-2}\ {}^{+22}_{-\ 4}\,\mathrm{GeV} \\
  M_{\mathrm D^*} &=&
    2.002^{+7}_{-6}\ {}^{+16}_{-41}\,\mathrm{GeV} \\
  M_{\mathrm D^*_{\mathrm s}} &=&
    2.082^{+4}_{-5}\ {}^{+22}_{-31}\,\mathrm{GeV} 
\end{array}
\qquad\qquad
\begin{array}{rcl}
\multicolumn{3}{c}{\mbox{Experiment~\cite{pdg2000}}}\\
  M_{\mathrm D^+_{\mathrm s}} &=& 1.9685^{+5}_{-5} \,\mathrm{GeV}\\
  M_{\mathrm D^{*0}} &=& 2.0067^{+5}_{-5}\,\mathrm{GeV} \\
  M_{\mathrm D^{*+}_{\mathrm s}} &=& 2.1124^{+7}_{-7}\,\mathrm{GeV}   
\end{array}
\end{equation}
For the lattice results, the first error is statistical and the second
systematic.  In this calculation the light quark in each meson is a
normal quark. Thus the $\mathrm D^*$ is the isospin-averaged vector
state, with mass $M_{\mathrm D^*} = (M_{\mathrm D^{*0}}+M_{\mathrm
D^{*\pm}})/2$.  With somewhat large systematic errors, the largest of
which comes from the lattice spacing, the spectrum is in broad
agreement with experiment.

\begin{table}
\begin{center}
\caption{The spectrum of heavy-light mesons with
        different definitions of the lattice spacing in GeV.\smallskip}
\label{tab:D_spec}
\begin{tabular}{cccc}\hline\hline
state&$a^{-1}$& $\beta=6.2$ &$\beta=6.0$ \\\hline
& $r_0$ & $1.956^{+2}_{-2}$ & $1.952^{+2}_{-2}$ \\
$\mathrm{D}_\mathrm{s}$ &$m_{\rho}$ & $1.963^{+2}_{-3}$ & $1.958^{+3}_{-3}$ \\
& $f_\pi$ & $1.960^{+4}_{-4}$ & $1.958^{+4}_{-3}$ \\\hline
& $r_0$ & $2.002^{+7}_{-6}$ & $2.011^{+8}_{-7}$ \\
$\mathrm{D}^*$ &$m_{\rho}$ & $1.960^{+6}_{-6}$ & $1.973^{+7}_{-6}$ \\
& $f_\pi$ & $1.973^{+10}_{-10}$ & $1.977^{+14}_{-14}$ \\\hline
& $r_0$ &  $2.082^{+4}_{-5}$ &  $2.083^{+4}_{-4}$\\
$\mathrm{D}^*_\mathrm{s}$ &$m_{\rho}$& $2.052^{+3}_{-4}$ & $2.053^{+4}_{-3}$\\
& $f_\pi$ & $2.061^{+7}_{-7}$ & $2.056^{+10}_{-10}$ \\\hline\hline
\end{tabular}
\end{center}
\end{table}

\begin{figure}
\begin{center}
\epsfig{file=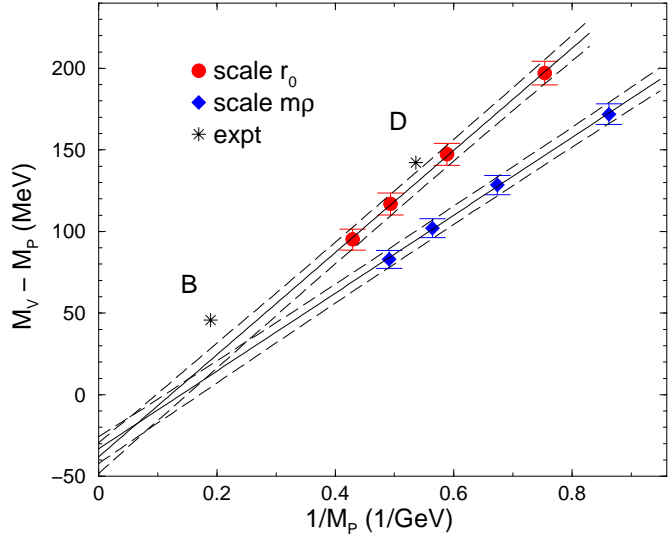,width=0.63\hsize} 
\caption{The hyperfine splitting as a function of $M$ in physical units at
        $\beta=6.2$. The circles have the scale set by $r_0$ and the 
        diamonds by $m_\rho$.}
\label{fig:splittings}
\end{center}
\end{figure}

It is also of interest to look at the $M_V-M_P$ splitting as a
function of the (pseudoscalar) meson mass. This can be extrapolated 
in inverse mass to the B scale, and to the infinite quark mass limit
where the splitting should vanish.  The splittings at both values of
$\beta$ are given in Table~\ref{tab:splittings} and are plotted for
$\beta=6.2$ in Figure~\ref{fig:splittings}. There is little difference
between linear and quadratic fits, so only the linear fit is
shown. The splittings obtained from the linear fit at $\beta=6.2$,
using $r_0$ to set the scale, are taken as the central values quoted
below. The second set of errors is systematic, obtained by combining
in quadrature the differences arising from the following variations in
procedure (in descending order of importance): using $m_\rho$ rather
than $r_0$ to set the scale; (for the B meson) using a quadratic
rather than linear extrapolation; multiple rather than single
exponential fits; using $\beta=6.0$ data rather than $\beta=6.2$
data. The results, with the comparison to experiment, are as follows:
\begin{equation}
\begin{array}{rcl}
\multicolumn{3}{c}{\mbox{Lattice, this work}}\\
  M_{\mathrm D^*}-M_{\mathrm D} &=&
       130^{+6}_{-6}{}^{+15}_{-35}\  \mathrm{MeV}\\
  M_{\mathrm B^*}-M_{\mathrm B} &=&
       \ 21^{+7}_{-8}{}^{+18}_{-16}\  \mathrm{MeV}
\end{array}
\quad\quad
\begin{array}{rcl}
\multicolumn{3}{c}{\mbox{Experiment~\cite{pdg2000}}}\\
   M_{\mathrm D^*}-M_{\mathrm D} &=& 142.6\pm0.5 \ \mathrm{MeV} \\
   M_{\mathrm B^*}-M_{\mathrm B} &=& \ 45.2\pm1.8 \ \mathrm{MeV}
\end{array}
\end{equation}
\begin{table}
\begin{center}
\caption{The hyperfine splitting of the heavy-light mesons at the B and
        D meson scale, in MeV.\smallskip}
\label{tab:splittings}
\begin{tabular}{cccc}\hline\hline
state & $a^{-1}$ &$\beta=6.2$ & $\beta=6.0$ \\\hline
& $r_0$ & $130^{+7}_{-6}$ & $140\ ^{+8}_{-6}$ \\
$M_{\mathrm D^*}-M_{\mathrm D}$ &$m_\rho$ & $95^{+6}_{-6}$& $108\ ^{+7}_{-6}$\\
& $f_\pi$ &$105^{+9}_{-9}$ & $111^{+14}_{-13}$ \\\hline
& $r_0$ & $21^{+7}_{-8}$ & $32\ ^{+8}_{-8}$ \\
$M_{\mathrm B^*}-M_{\mathrm B}$ & $m_\rho$& $12^{+6}_{-7}$& $22\ ^{+8}_{-8}$ \\
& $f_\pi$ &$15^{+8}_{-8}$ & $24^{+14}_{-14}$ \\\hline\hline
\end{tabular}
\end{center}
\end{table}

\section{Conclusions}
The decay constants of heavy-light and light mesons have been
determined in the quenched approximation at two values of the
coupling. The action and currents have been fully NP $\order{a}$
improved. Good scaling of the decay constants is found.

Uncertainties in the improvement and renormalisation coefficients
are seen to have a large effect on the decay constants. The ALPHA
and \Bhatt determinations of the improvement coefficients produce
different values of $f_\mathrm{K}$, especially at $\beta=6.0$.
However, the $f_\mathrm{K}$ do converge as the lattice spacing
decreases.  Computed quantities may well differ at $\order{a^2}$ at
fixed lattice spacing but should agree in the continuum limit.

The heavy-light decay constants are found to be around $10\%$ higher
than, but compatible with, the world average~\cite{Bernard_2000} of
quenched lattice determinations.  HQS relations are found to be well
satisfied by the data at both values of the coupling. The systematic
uncertainties in the calculation have been discussed and, with the
exception of quenching, estimated.  The issue of $\order{a^2m^2}$
effects in the heavy quark extrapolations has been considered in
various ways. In particular, the difference of the decay constants
between $\beta=6.0$ and $\beta=6.2$ at fixed heavy quark mass, as a
measure of discretisation effects, is small and remains so during the
heavy quark extrapolation. The static point value of the decay
constant, despite the lack of improvement, suggests the extrapolation
is under control.  Furthermore, HQS relations are satisfied and there
is agreement between $f_\mathrm{B_s}/f_\mathrm{B}$ computed by
extrapolating the ratio and by taking the ratio of the extrapolated
decay constants.

The quenched spectrum of D mesons has been determined, shows good
scaling, and is found to be in broad agreement with experimental data.

\begin{ack}
Support from EPSRC grant GR/K41663, and PPARC grants GR/L29927,
GR/L56336 and PPA/G/S/1999/00022 is acknowledged.  DGR acknowledges
PPARC and the DOE (contract DE-AC05-84ER40150), and thanks FNAL for
hospitality during part of this work. LDD thanks MURST for financial
help. CMM acknowledges PPARC grant PPA/P/S/1998/00255. GNL thanks the
Edinburgh Department of Physics and Astronomy for financial support.
The authors thank Sara Collins, Christine Davies, Martin
Kurth, Craig McNeile, Sin\'{e}ad Ryan and Hartmut Wittig for useful
discussions.
\end{ack}

\end{document}